\documentstyle[12pt,axodraw]{article}
\begin{document}
\baselineskip 0.6cm

\def\simgt{\mathrel{\lower2.5pt\vbox{\lineskip=0pt\baselineskip=0pt
           \hbox{$>$}\hbox{$\sim$}}}}
\def\simlt{\mathrel{\lower2.5pt\vbox{\lineskip=0pt\baselineskip=0pt
           \hbox{$<$}\hbox{$\sim$}}}}

\begin{titlepage}

\begin{flushright}
UCB-PTH-02/55 \\
LBNL-51782 \\
FERMILAB-Conf-02/322-T
\end{flushright}

\vskip 1.0cm

\begin{center}
{\Large \bf Grand Unification in Higher Dimensions}

\vskip 0.8cm

{\large
Lawrence J.~Hall$^{a,b}$ and Yasunori Nomura$^c$
}

\vskip 0.4cm

$^a$ {\it Department of Physics, University of California,
                Berkeley, CA 94720}\\
$^b$ {\it Theoretical Physics Group, Lawrence Berkeley National Laboratory,
                Berkeley, CA 94720}\\
$^c$ {\it Theoretical Physics Department, Fermi National Accelerator 
                Laboratory, Batavia, IL 60510}

\vskip 1.2cm

\abstract{
We have recently proposed an alternative picture for the physics at 
the scale of gauge coupling unification, where the unified symmetry 
is realized in higher dimensions but is broken locally by a symmetry 
breaking defect. Gauge coupling unification, the quantum numbers of 
quarks and leptons and the longevity of the proton arise as phenomena 
of the symmetrical bulk, while the lightness of the Higgs doublets 
and the masses of the light quarks and leptons probe the symmetry 
breaking defect. Moreover, the framework is extremely predictive 
if the effective higher dimensional theory is valid over a large 
energy interval up to the scale of strong coupling. Precise agreement 
with experiments is obtained in the simplest theory --- $SU(5)$ in 
five dimensions with two Higgs multiplets propagating in the bulk. 
The weak mixing angle is predicted to be $\sin^2\theta_w = 0.2313 
\pm 0.0004$, which fits the data with extraordinary accuracy. The 
compactification scale and the strong coupling scale are determined 
to be $M_c \simeq 5 \times 10^{14}~{\rm GeV}$ and $M_s \simeq 1 
\times 10^{17}~{\rm GeV}$, respectively. Proton decay with a lifetime 
of order $10^{34}~{\rm years}$ is expected with a variety of final 
states such as $e^+\pi^0$, and several aspects of flavor, including 
large neutrino mixing angles, are understood by the geometrical 
locations of the matter fields. When combined with a particular 
supersymmetry breaking mechanism, the theory predicts large lepton 
flavor violating $\mu \rightarrow e$ and $\tau \rightarrow \mu$ 
transitions, with all superpartner masses determined by only two 
free parameters. The predicted value of the bottom quark mass from 
Yukawa unification agrees well with the data. This paper is mainly 
a review of the work presented in hep-ph/0103125, hep-ph/0111068 
and hep-ph/0205067~\cite{Hall:2001pg, Hall:2001xb, Hall:2002ci}.
}

\end{center}
\end{titlepage}

\section{Introduction: Features of 4D Grand Unification}
\label{sec:4dguts}

While the manifestations of the strong, weak and electromagnetic forces 
are very different in nature, these three interactions all follow from
local gauge symmetry, suggesting that they may be low energy remnants 
of a single large gauge symmetry at high energies. Such a grand unified 
interaction would be described by a single gauge coupling, leading to a 
correlation among the strengths of the three forces measured at lower 
energies. Remarkably, this prediction from gauge coupling unification 
is highly successful, if nature is supersymmetric above the scale of 
the weak interactions. It implies that the unification of the strong, 
and electroweak interactions occurs at a mass scale of order 
$10^{16}~{\rm GeV}$, but what is the nature of the physical theory 
underlying this unification and how can it be experimentally tested?

The conventional answer of four dimensional (4D) grand 
unification~\cite{Georgi:sy} shows a remarkable dichotomy: parts of 
the standard model cry out for 4D unification into a gauge group such 
as $SU(5)$, while other parts abhor such a unification. For example 
the quantum numbers of a generation of quarks and leptons fit 
beautifully into unified representations, providing an elegant 
understanding of the various gauge quantum numbers~\cite{Pati:1974yy, 
Georgi:sy, SO10}, while the Higgs doublet resists unification. 
The $SU(5)$ partner of the Higgs doublet, $H_3$, must be heavy to
avoid rapid proton decay and also because it would spoil gauge
coupling unification. Whilst the simplest picture of supersymmetric
grand unification gives us a very significant prediction for the weak
mixing angle~\cite{Dimopoulos:1981zb, Dimopoulos:1981yj}, it also leads 
to a prediction for proton decay from the exchange of the superheavy 
triplets $H_3$~\cite{Sakai:1981pk}, in strong disagreement with
data~\cite{Goto:1998qg}. Finally, the mass ratio of quarks and leptons 
in the third generation, $m_b/m_\tau$, shows a simple ratio which 
follows directly from grand unification~\cite{Chanowitz:1977ye}, while 
light quark-lepton mass ratios, such as $m_s/m_\mu$, do not have 
values that follow simply from unification. Thus the minimal theory 
does not explain why there is a light Higgs boson, is excluded by 
proton decay, and introduces flavor conundrums. Of course, this 
dichotomy does not exclude supersymmetric 4D unification which has 
been so much discussed for over 20 years; rather, within these theories 
we are led to invent a series of mechanisms for doublet-triplet 
splitting, proton decay suppression, and flavor. However, the resulting 
theories then acquire a certain level of complexity. Can the dichotomy 
be resolved more elegantly in an alternative framework?

\begin{figure}
\begin{center} 
\begin{picture}(275,300)(-65,0)
  \LongArrow(0,0)(0,300) \Text(-5,300)[r]{$\alpha_s(M_Z)$}
  \Line(-3,15)(3,15)   \Text(-8,16)[r]{$0.060$}
  \Line(-3,45)(3,45)
  \Line(-3,75)(3,75)   \Text(-8,76)[r]{$0.080$}
  \Line(-3,105)(3,105)
  \Line(-3,135)(3,135) \Text(-8,136)[r]{$0.100$}
  \Line(-3,165)(3,165)
  \Line(-3,195)(3,195) \Text(-8,196)[r]{$0.120$}
  \Line(-3,225)(3,225)
  \Line(-3,255)(3,255) \Text(-8,256)[r]{$0.140$}
  \Line(-3,285)(3,285)
% experiment
  \Text(160,186)[l]{$\alpha_s^{\rm exp}$}
  \DashLine(0,180)(150,180){4}
  \DashLine(0,192)(150,192){4}
% GUT
  \Text(50,285)[b]{$\alpha_s^{\rm GUT}$}
  \DashLine(50,60)(50,72){1} \Vertex(50,66){3}
  \DashLine(45,60)(55,60){1} \DashLine(45,72)(55,72){1} 
% SUSY GUT
  \Text(100,285)[b]{$\alpha_s^{\rm SGUT}$}
  \DashLine(100,216)(100,234){1}
  \DashLine(95,216)(105,216){1} \DashLine(95,234)(105,234){1}
  \Line(100,213)(100,237) \Vertex(100,225){3}
  \Line(95,213)(105,213) \Line(95,237)(105,237)
  \LongArrow(53,80)(93,206) \Text(75,140)[tl]{SUSY log}
\end{picture}
\caption{The predictions for $\alpha_s(M_Z)$ in non-supersymmetric 
 grand unification, $\alpha_s^{\rm GUT}$, and supersymmetric grand 
 unification, $\alpha_s^{\rm SGUT}$. The solid error bar represents 
 the threshold corrections from the superpartner spectrum. Dotted 
 error bars represent threshold corrections from the unified scale 
 corresponding to a heavy ${\bf 5} + \bar{\bf 5}$ representation 
 with unit logarithmic mass splitting between doublets and triplets.} 
 \label{fig:alphas_1}
\end{center}
\end{figure}
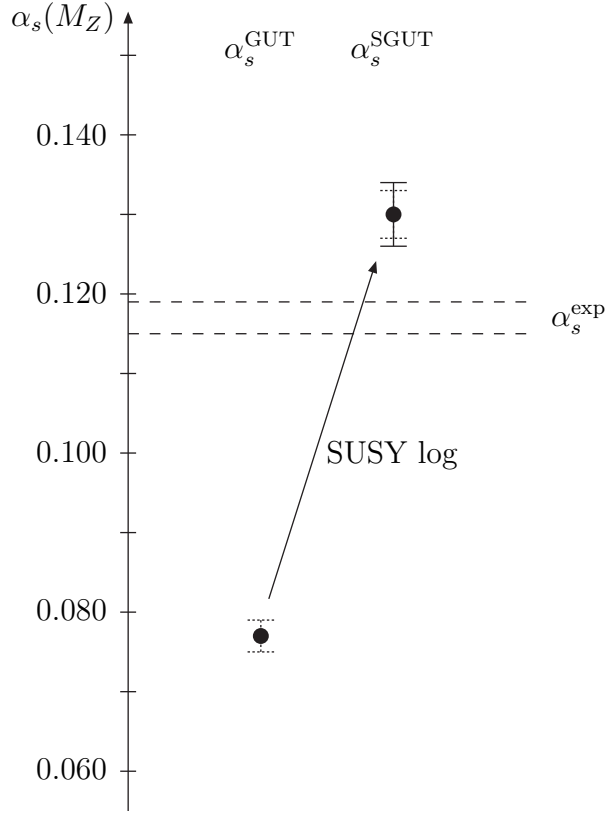
The prediction of the QCD gauge coupling from 4D supersymmetric
unification is good but certainly not perfect, as illustrated in
Figure~\ref{fig:alphas_1}. The effect of the supersymmetric 
logarithm is to greatly improve the prediction, but there is 
an overshoot beyond the experimental value of $\alpha_s(M_Z) 
= 0.117 \pm 0.002$~\cite{Hagiwara:pw} to $\alpha_s(M_Z) \simeq 
0.130$~\cite{Langacker:1995fk}. We typically assume that most of 
this discrepancy comes from the unification scale, and is due to 
the complications to the theory that we have been forced to add. 
Since these corrections involve additional free parameters, they 
cannot be numerically evaluated. In this talk we will argue that 
there is an alternative picture for the physics at the unification 
scale, and that all aspects of the dichotomy are reconciled in the 
simplest model.  In the next section we introduce the new picture 
of symmetry breaking defects in a higher dimensional spacetime, 
and argue that the traditional problems are all elegantly solved. 
In section~\ref{sec:minmodel} we discuss the predictions from gauge 
coupling and Yukawa coupling unification in the simplest model. 
The corrections to the QCD gauge coupling from physics at the unified 
scale do not depend on any extra free parameters, and yield precisely 
the observed value. The location of matter in the higher dimension 
is discussed, and predictions for proton decay are given. 
In section~\ref{sec:expsig} we introduce a new origin for 
supersymmetry breaking in unified theories.  Combining this with 
the minimal model, predictions are given for the superpartner and 
Higgs spectrum, for the bottom quark mass from Yukawa unification, 
and for flavor changing lepton decays. We conclude in 
section~\ref{sec:conc}.

\section{New Physics for Grand Unification}
\label{sec:newphysics}

Building on the ideas and tools developed by others, over the last
year or so we have introduced a new picture for the physics in the
energy range of $10^{15}$--$10^{17}~{\rm GeV}$~\cite{Hall:2001pg, 
Hall:2001xb, Hall:2002ci}. We describe the new physical picture 
in this section, and the simplest model for its implementation 
in the next.

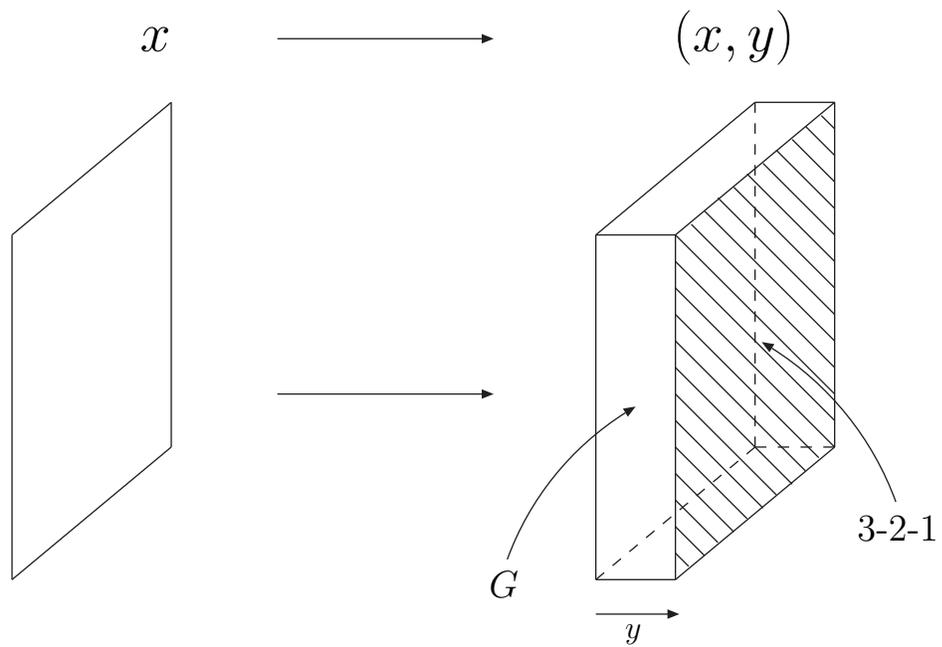
\begin{figure}
\begin{center} 
\begin{picture}(350,240)(-20,-30)
  \Line(0,0)(0,130) \Line(60,50)(60,180)
  \Line(0,130)(60,180) \Line(0,0)(60,50)
  \LongArrow(100,70)(180,70)
  \Text(60,200)[rb]{\LARGE $x$}
  \LongArrow(100,204)(180,204)
  \Text(250,200)[lb]{\LARGE $(x, y)$}
  \DashLine(280,180)(280,50){4}
  \DashLine(280,50)(310,50){4}
  \DashLine(280,50)(220,0){4}
  \Line(310,170)(305,175) \Line(310,160)(299,171)
  \Line(310,150)(294,166) \Line(310,140)(288,162)
  \Line(310,130)(283,157) \Line(310,120)(277,153)
  \Line(310,110)(272,148) \Line(310,100)(266,144)
  \Line(310,90)(261,139)  \Line(310,80)(256,134)
  \Line(310,70)(250,130)  \Line(310,60)(250,120)
  \Line(250,110)(310,50) \Line(250,100)(304,46)
  \Line(250,90)(299,41)  \Line(250,80)(294,36)
  \Line(250,70)(288,32)  \Line(250,60)(283,27)
  \Line(250,50)(277,23)  \Line(250,40)(272,18)
  \Line(250,30)(266,14)  \Line(250,20)(261,9)
  \Line(250,10)(255,5)
  \Line(220,0)(220,130) \Line(220,130)(280,180)
  \Line(220,0)(250,0) \Line(220,130)(250,130) \Line(280,180)(310,180)
  \Line(250,0)(250,130) \Line(310,50)(310,180)
  \Line(250,130)(310,180) \Line(250,0)(310,50)
  \LongArrow(220,-13)(250,-13) \Text(235,-17)[t]{$y$}
  \LongArrowArcn(290,-30)(110,160,120) \Text(186,3)[t]{\large $G$}
  \LongArrowArc(225,-10)(115,20,60) \Text(335,24)[t]{\large 3-2-1}
\end{picture}
\caption{Physics at high energies probes extra dimensions, $y$, 
 that extend over small sizes $R$. The interactions in the volume 
 of this higher dimensional box are constrained by a unified gauge 
 symmetry $G$, but interactions on a boundary may be constrained 
 only by a smaller symmetry, such as 3-2-1, creating a defect of 
 lower dimension.}
\label{fig:picture}
\end{center}
\end{figure}

At the TeV scale we live in a 4D world, spanned by the coordinates
$x$, and the gauge group is $SU(3)_C \times SU(2)_L \times U(1)_Y$ 
(3-2-1), as illustrated by the sheet on the left-hand side of 
Figure~\ref{fig:picture}. At the unification scale we suppose that 
other dimensions of size $R$ are resolved, described by coordinates 
$y$, as shown on the right of the figure. It is the mass scale $1/R$, 
rather than the expectation value of some field, that characterizes 
the scale of unification. Particles moving in the $y$ direction can 
be viewed as particles moving in a box of size $R$ and therefore 
have momenta $p_y$ quantized in units of $1/R$. To observers in 4D, 
particles with different $p_y$ appear as particles of different mass, 
so that there is a discrete tower of particles, known as the 
Kaluza-Klein (KK) tower. A crucial aspect of our physical picture 
is the structure of the gauge symmetries in the box of the $y$ 
direction. Interactions in the interior of the box are symmetrical 
under the full gauge symmetry $G$ of the unified theory, while those 
on a boundary are only symmetrical under the standard model 3-2-1 
gauge symmetries, as shown in Figure~\ref{fig:picture}.

\begin{figure}
\begin{center} 
\begin{picture}(250,150)(-5,-30)
  \DashLine(0,0)(100,50){4} \DashLine(100,50)(230,50){4} 
  \DashLine(100,50)(100,120){4}
  \Line(0,0)(130,0) \Line(0,70)(130,70)
  \Line(0,0)(0,70) \Line(130,0)(130,70)
  \Line(0,70)(100,120) \Line(100,120)(230,120)
  \Line(130,70)(230,120) \Line(230,50)(230,120)
  \LinAxis(130,0)(230,50)(1,1,1,1,2)
  \LongArrow(0,-13)(30,-13) \Text(15,-17)[t]{$y_1$}
  \LongArrow(-13,0)(-13,30) \Text(-17,15)[r]{$y_2$}
  \LongArrow(7,14)(33,27) \Text(17,27)[b]{$y_3$}
  \LongArrowArc(167,-11)(40,20,60) \Text(206,0)[t]{\large 3-2-1}
  \Text(130,105)[t]{\LARGE $G$}
\end{picture}
\caption{An example of a 3D bulk with a 1D defect.}
\label{fig:extra-3D}
\end{center}
\end{figure}
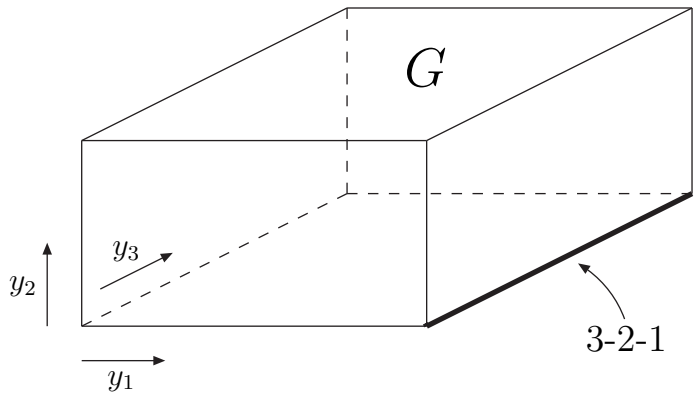

Since our own four dimensions are known, it is convenient to suppress 
$x$ and display only the extra dimensions $y$, as illustrated 
in Figure~\ref{fig:extra-3D} for the case of three extra dimensions. 
The space of the extra dimensions is known as the bulk, and we will 
also refer to it as a box. The sizes of the extra dimensions need not
be the same, although we imagine they are not extremely different, 
and, while we have shown a simple box, the bulk may have a more
complicated geometry. The crucial point is that the gauge group
throughout the volume of this extra-dimensional bulk is the unified
group $G$, while that of the standard model appears only on some lower
dimensional boundary surface. In Figure~\ref{fig:extra-3D} we have shown 
the 3-2-1 surface as having dimension 1, but it could be a 2D surface, or 
it could be a point at one of the corners of the box. In our picture,
most of spacetime feels the full gauge invariance of $G$, while 
there is a defect on a lower dimensional surface which only feels 
the 3-2-1 gauge symmetry. The symmetry breaking therefore appears 
explicitly, as a spatial defect --- a complete change of viewpoint
compared to 4D grand unification! One might naively guess that such
local defects could not lead to the world we see -- where 3-2-1 forces
are observed to be quite different from each other, and the other 
interactions in $G$ are incredibly feeble. Figure~\ref{fig:extra-3D} 
gives the impression that the breaking of $G$ is minor and perhaps 
just a small correction. For short distance physics in the bulk this 
is certainly true -- but for long distance physics the boundary 
effects become all important, as they determine the light states 
of the theory.

How does this new picture reconcile the dichotomy of 4D unification? 
The idea is remarkably simple: the aspects which fit unification so 
well, gauge coupling unification, quark and lepton quantum numbers and 
$m_b/m_\tau$, should be phenomena of the regions of the bulk where 
the physics is unified and should be insensitive to the 3-2-1 defect, 
while the aspects which abhor unification, such as the light Higgs 
doublet and $m_s/m_\mu$, should probe the 3-2-1 defect in a non-trivial
way. It will turn out that the suppression of proton decay results
from the enlargement of the spacetime symmetry of the bulk.

\begin{figure}
\begin{center}
  \input{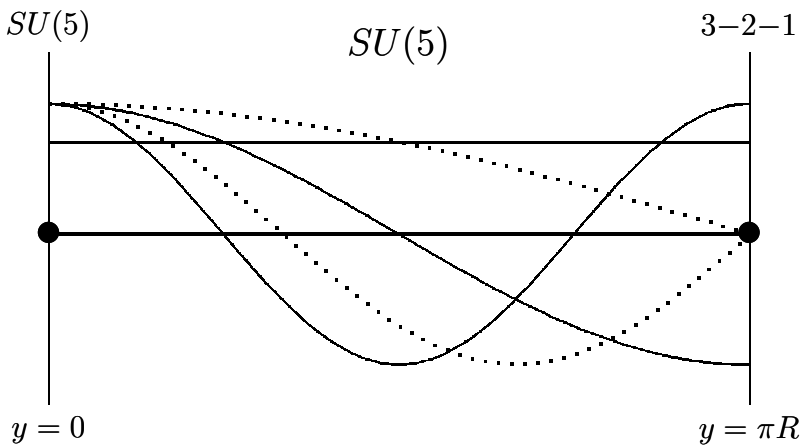}
\caption{In the fifth dimension, space is a line segment with 
 boundaries at $y=0$ and at $y=\pi R$.  Solid and dotted lines 
 represent the profiles of gauge transformation parameters $\xi_{321}$ 
 and $\xi_X$, respectively.  Because $\xi_X(y = \pi R) = 0$, 
 a point defect occurs in the symmetry at the $y = \pi R$ boundary, 
 explicitly breaking $SU(5)$ to $SU(3)_C \times SU(2)_L \times U(1)_Y$.}
\label{fig:orbifold}
\end{center}
\end{figure}

What is the origin of the 3-2-1 defect? In particular what determines
its location and why is it on a boundary of the bulk rather than
somewhere in the interior? We will answer this in the context of the
effective higher dimensional field theory. We do not explain the
origin of the extra dimensions nor their size, but, given this enlarged 
spacetime, we write the most general theory subject to a set of
symmetries. The new ingredient here, compared with familiar 4D 
theories, is that because our spacetime has boundaries we must specify 
boundary conditions to define the theory.  The boundary conditions are 
chosen not to spoil the consistency of the theory, and, within the 
effective field theory description, the 3-2-1 defect originates from 
these boundary conditions. Specifically, in 4D theories we take the 
gauge parameters for transformation $a$ to be arbitrary functions of 
spacetime, $\xi_a(x)$, but with a finite bulk we must specify boundary 
conditions for these parameters. In particular, at some boundary 
$y=y_b$ we can specify different conditions on the 3-2-1 gauge 
parameters, $\xi_{321}(y_b)$, and the remaining gauge parameters 
of $G$, $\xi_X(y_b)$, inducing the $G$-breaking defect on this 
boundary. For example, if $\xi_X(y_b) = 0$ and $\xi_{321}(y_b) \neq 0$ 
then the unified gauge bosons $X$ do not have interactions on this 
boundary, while the 3-2-1 gauge bosons of the standard model do. 
This is illustrated for the case $G=SU(5)$ in a 1D bulk in 
Figure~\ref{fig:orbifold}. In this example the fields and interactions 
at the $y = \pi R$ boundary need only respect 3-2-1 gauge symmetry --- 
they explicitly break the $SU(5)$ symmetry~\cite{Hall:2001pg}.

This is clearly a radical departure from the familiar Higgs 
mechanism for spontaneously breaking gauge symmetries. There is no 
Higgs field --- in the effective field theory the phenomena are 
geometrical rather than dynamical. In cosmology, as the temperature 
$T$ of the universe cools through $1/R$, there is no phase transition; 
rather the symmetry breaking effects gradually grow in importance. 
At $T \gg 1/R$ they are important only very close to the defect, and 
irrelevant everywhere else. At lower temperatures they become ever 
more dominant, and the symmetries of the bulk cannot be resolved.
Remarkably, this explicit breaking of gauge symmetry does not destroy 
calculability of the theory. The unitarity behavior of the theory 
is no worse than when the boundary conditions preserve 
$SU(5)$~\cite{Hall:2001tn}.

\begin{figure}
\begin{center} 
\begin{picture}(420,170)(0,-30)
  \DashLine(0,0)(70,20){4} \DashLine(70,20)(180,20){4} 
  \DashLine(70,20)(70,90){4}
  \Line(0,0)(110,0) \Line(0,70)(110,70)
  \Line(0,0)(0,70) \Line(110,0)(110,70)
  \Line(0,70)(70,90) \Line(70,90)(180,90)
  \Line(110,70)(180,90) \Line(180,20)(180,90)
  \LinAxis(110,0)(180,20)(1,1,1,1,2) \Text(147,8)[tl]{\large 3-2-1}
  \Text(55,32)[b]{\large $\phi(y)$}
  \Line(215,42)(260,42) \Line(262,40)(256,46)
  \Line(215,38)(260,38) \Line(262,40)(256,34)
  \Line(310,-20)(420,-20) 
  \LongArrow(310,-20)(310,140) \Text(305,140)[r]{$m$}
  \LinAxis(340,-20)(380,-20)(1,1,1,1,2)
  \LinAxis(340,30)(380,30)(1,1,1,1,2)
  \LinAxis(340,80)(380,80)(1,1,1,1,2)
  \LinAxis(340,130)(380,130)(1,1,1,1,2)
  \Line(308,30)(312,30) \Text(305,30)[r]{$1/R$}
\end{picture}
\caption{A bulk, having boundary conditions leading to a defect, can be 
 viewed as a ``machine'' for creating a mass spectrum of 4D particles,
 known as a KK tower.}
\label{fig:machine}
\end{center}
\end{figure}
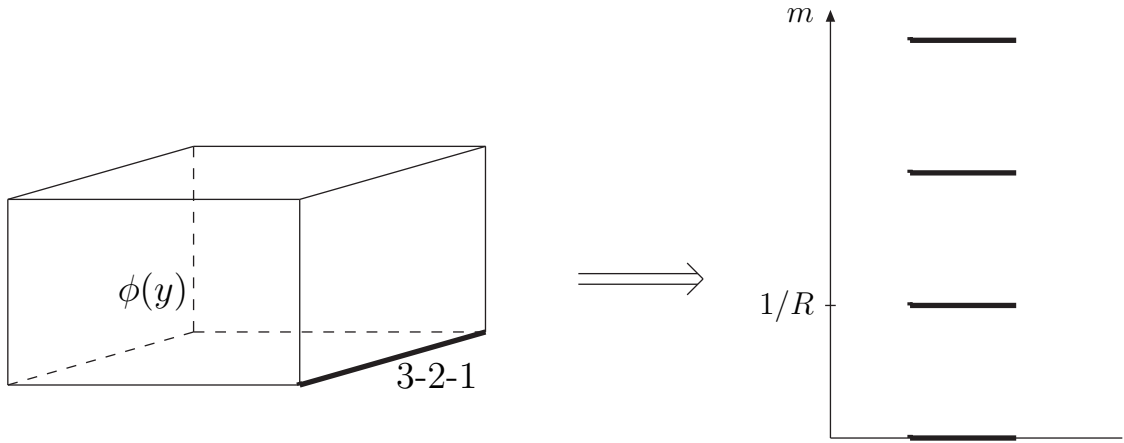

What happens when fields $\phi(x,y)$ propagate in such a higher 
dimensional spacetime with a gauge symmetry defect induced by 
non-trivial boundary conditions? The geometry of the box together with
the boundary conditions determine the allowed normal modes. This is
just the field theory analogue of quantizing a particle in a box, but
now the allowed  $p_y^2$ correspond to the allowed $m^2$ for a 4D
observer. Thus the box and boundary conditions can be viewed as a
machine for creating a KK tower of massive states, as illustrated in
Figure~\ref{fig:machine}.

The typical spacing or discreteness, $1/R$, for the masses of the KK 
tower is determined by the size of the box, while the gauge quantum 
numbers at each level is determined by the boundary conditions. 
The important point is that modes having different 3-2-1 quantum 
numbers $q_1,q_2,\cdots$, have different spectra, because of the 
$G$-breaking boundary conditions. At high energies the discreteness 
is not so important and the local symmetry $G$ of the bulk is restored, 
but at low energies the discreteness is crucial and the gauge quantum 
numbers of the lowest lying states are all important. At very low 
energies, only the zero mass modes can be excited, and hence a crucial 
question is how the boundary conditions determine the quantum numbers 
of these ``zero modes''. We can turn this around: at the scale of
unification we want only the particles of the standard model and 
their superpartners to be zero modes, so that running the machine 
backwards we can find out what geometries and boundary conditions 
are of interest for nature. The remarkable thing is that we can start 
with a pure gauge theory in the box --- there are no mass terms or 
spontaneous symmetry breakings --- and the 4D mass terms of the KK 
tower arise from the kinetic energies in the extra dimensions. 
What are the consequences of this new viewpoint for gauge coupling 
unification, the lightness of the Higgs doublet, proton decay, and 
quark-lepton mass ratios? We will not be surprised to see large 
changes from the standard picture.

\subsection{Gauge coupling unification}

Since the unified symmetry $G$ is explicitly broken by boundary 
conditions, it is not obvious that gauge coupling unification is 
preserved. In fact, gauge coupling unification is generically 
destroyed due to the presence of local $G$ breaking on the 
$y = y_b$ boundary.  To see this, consider the effective field 
theory above $1/R$.  Since the higher dimensional gauge theory is 
non-renormalizable, this effective theory must be cut off at some 
scale $M_s$, where the theory is embedded into a more fundamental 
theory such as string theory.  At the scale $M_s$, the most general 
effective action for the gauge kinetic terms is
\begin{equation}
  S = \int dx \; dy \; 
    \biggl[ \frac{1}{g_5^2} F^2 + 
    \delta(y - y_b) \frac{1}{\tilde{g}_a^2} F_a^2 \biggr],
\label{eq:gaugekinops}
\end{equation}
where the first term arises from the interior of the box and is $G$ 
invariant, while the second term represents non-unified kinetic 
operators located on the $y=y_b$ boundary ($F$ is the field strength, 
and $a = 1,2,3$ represents the standard model gauge groups).  This form 
is ensured by the $y$-dependent gauge symmetry of our effective theory, 
regardless of the unknown ultraviolet physics above $M_s$. The standard 
model gauge couplings in the equivalent 4D theory, $g_a$, are then 
obtained by integrating over the extra dimensions:
\begin{equation}
  \frac{1}{g_a^2} = \frac{R^d}{g_5^2} + \frac{R^{d'}}{\tilde{g}_a^2},
\label{eq:4d-gi}
\end{equation}
for $d$ extra dimensions of size $R$ and defects of dimension $d'<d$.
This shows that $g_a$ depend on the coefficients of the localized 
kinetic operators, $\tilde{g}_a$, and are not universal at the scale 
$M_s$ --- {\it in general there is no gauge coupling unification!} 
However, if the extra dimensions have a large size, we find that the 
$R^d$ factor dominates over $R^{d'}$, ensuring the unified contribution 
dominates over the 3-2-1 defect contribution, and gauge coupling 
unification is recovered~\cite{Hall:2001pg}.

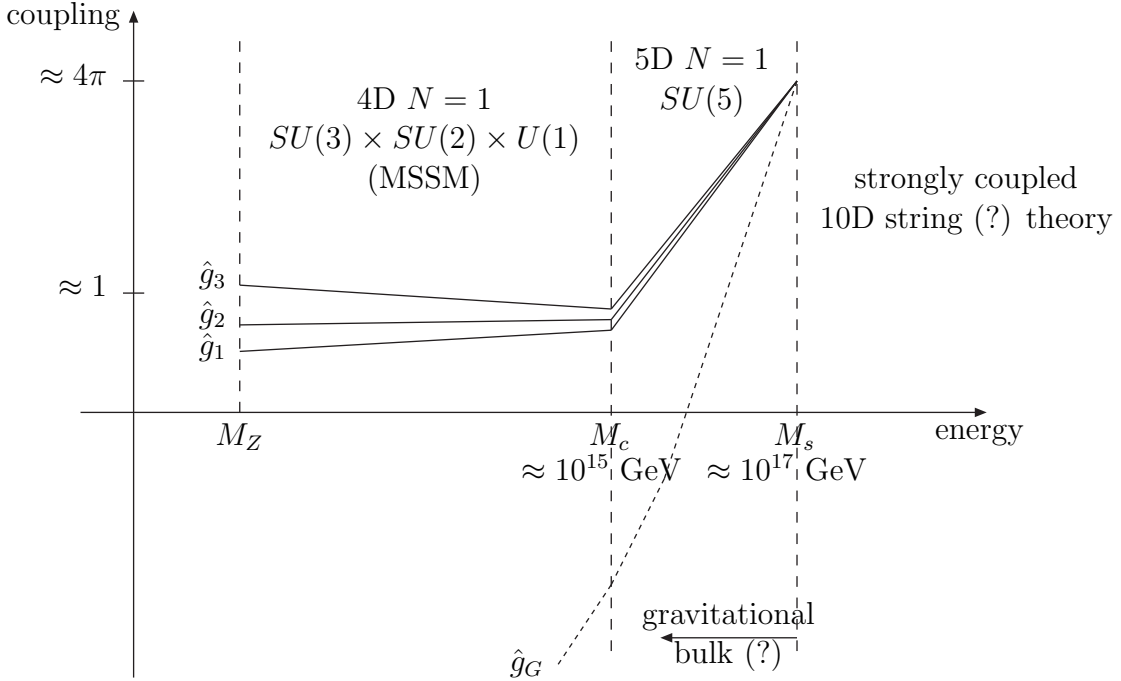
\begin{figure}
\begin{center} 
\begin{picture}(350,250)(-10,-100)
  \LongArrow(0,-100)(0,150) \Text(-5,150)[r]{coupling}
  \Line(-4,125)(4,125) \Text(-10,128)[r]{$\approx 4\pi$}
  \Line(-4,45)(4,45) \Text(-10,48)[r]{$\approx 1$}
  \LongArrow(-20,0)(320,0) \Text(320,-5)[t]{energy}
  \DashLine(40,0)(40,140){4} \Text(40,-4)[t]{$M_Z$}
  \DashLine(180,-90)(180,140){4} 
  \Text(180,-4)[t]{$M_c$} \Text(177,-16)[t]{$\approx 10^{15}~{\rm GeV}$}
  \DashLine(250,-90)(250,140){4}
  \Text(250,-4)[t]{$M_s$} \Text(247,-16)[t]{$\approx 10^{17}~{\rm GeV}$}
% standard model gauge couplings
  \Line(250,125)(180,39) \Line(180,39)(40,48)
  \Text(35,52)[r]{$\hat{g}_3$}
  \Line(250,125)(180,35) \Line(180,35)(40,33)
  \Text(35,37)[r]{$\hat{g}_2$}
  \Line(250,125)(180,31) \Line(180,31)(40,23)
  \Text(35,25)[r]{$\hat{g}_1$}
% gravitational couplings
  \DashLine(250,125)(200,-25){2}
  \DashLine(200,-25)(180,-65){2}
  \DashLine(180,-65)(160,-95){2}
  \Text(155,-95)[r]{$\hat{g}_G$}
% theories
  \Text(110,115)[b]{4D $N=1$}
  \Text(110,100)[b]{$SU(3) \times SU(2) \times U(1)$}
  \Text(110,85)[b]{(MSSM)}
  \Text(215,130)[b]{5D $N=1$}
  \Text(215,115)[b]{$SU(5)$}
  \Text(315,85)[b]{strongly coupled}
  \Text(315,70)[b]{10D string (?) theory}
  \LongArrow(250,-85)(200,-85)
  \Text(225,-80)[b]{gravitational} \Text(225,-86)[t]{bulk (?)}
\end{picture}
\caption{The energy dependence of the strengths of the gauge interactions.}
\label{fig:frame}
\end{center}
\end{figure}
The energy dependence of the gauge couplings in our scheme is shown in 
Figure~\ref{fig:frame}. The estimate for the compactification scale, 
$M_c \equiv 1/R \approx 10^{15}~{\rm GeV}$, and for the fundamental 
scale, $M_s \approx 10^{17}~{\rm GeV}$, are for the particular 5D theory 
with $G = SU(5)$ discussed in the next section, but the general behavior 
of the running of these couplings is generic to our framework. The 
couplings do not unify at the compactification scale; rather they 
continue to evolve even above the compactification scale where the 
physics is higher dimensional. The higher dimensional behavior of the 
theory is apparent because of the rapid growth of the interaction 
strength with energy. Above $M_c$ the couplings continue to approach 
each other because of the $G$-violating effects of the 3-2-1 defect.
Unification finally occurs at the fundamental scale. It is well known
that a unification with two mass scales, such as $M_c$ and $M_s$,
leads to a loss of predictivity of the low energy gauge couplings,
since they depend on the extra parameter $M_s/M_c$. We overcome this
by assuming that the theory at $M_s$ is strongly coupled, so that this
mass ratio is predicted~\cite{Hall:2001xb}:
\begin{equation}
  \left( {M_s \over M_c} \right)^d \approx {16 \pi^2 \over C g^2(M_c)},
\label{eq:strongcoup}
\end{equation}
where $C$ is a group theory Casimir; for example $C \approx 5$ for 
$G = SU(5)$. Given the rapid growth in the gauge couplings above $M_c$,
and the strong coupling of the theory at $M_s$, one may wonder whether
the unification can be reliably computed. Remarkably, however, the 
framework turns out to be extremely predictive.  The strong coupling 
requirement allows us to reliably estimate the size of non-unified 
corrections from unknown ultraviolet physics, and the precise prediction 
for the low energy QCD coupling is obtained as long as the volume of 
the 3-2-1 defects is sufficiently small~\cite{Hall:2001xb, Nomura:2001tn}.
The uncertainties in the estimate of Eq.~(\ref{eq:strongcoup}), for 
example from power-law corrections to gauge couplings~\cite{Dienes:1998vh}, 
are also well under control and have little effect on the 
prediction~\cite{Hall:2001xb}.

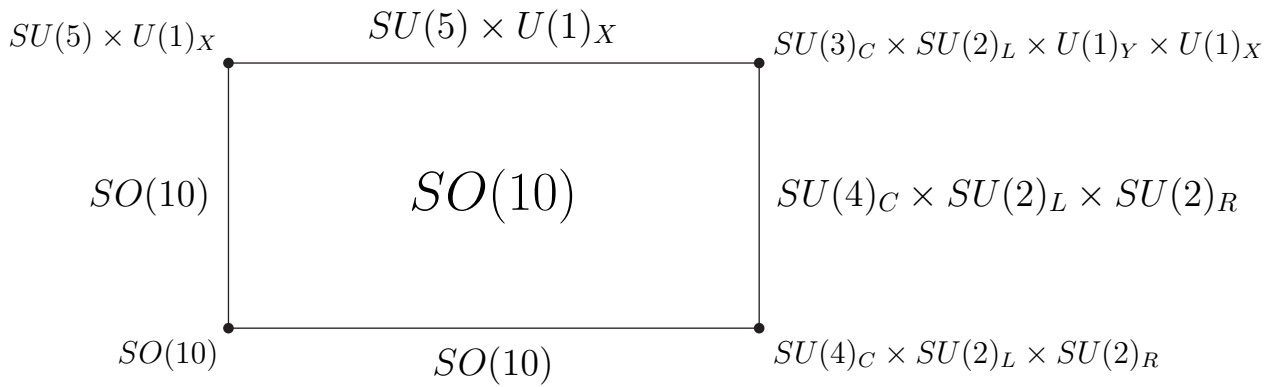
\begin{figure}
\begin{center} 
\begin{picture}(350,145)(-30,-25)
  \Text(100,50)[]{\LARGE $SO(10)$}
  \Line(0,0)(200,0) \Text(100,-7)[t]{\large $SO(10)$}
  \Line(0,0)(0,100) \Text(-7,50)[r]{\large $SO(10)$}
  \Line(0,100)(200,100) \Text(100,110)[b]{\large $SU(5) \times U(1)_X$}
  \Line(200,0)(200,100) \Text(207,50)[l]{\large $SU(4)_C
                          \times SU(2)_L \times SU(2)_R$}
  \Vertex(0,0){2} \Text(-4,-4)[tr]{$SO(10)$}
  \Vertex(0,100){2} \Text(-4,107)[br]{$SU(5) \times U(1)_X$}
  \Vertex(200,0){2} \Text(206,-4)[tl]{$SU(4)_C \times SU(2)_L \times SU(2)_R$}
  \Vertex(200,100){2} \Text(206,104)[bl]{$SU(3)_C \times SU(2)_L 
                        \times U(1)_Y \times U(1)_X$}
\end{picture}
\caption{A 2D bulk with $SO(10)$ gauge symmetry allows for an 
interesting set of defects. There are two 1D defects, one having 
the usual $SU(5)$ unified gauge symmetry and the other having the 
left-right $SU(4)_C \times SU(2)_L \times SU(2)_R$ symmetry introduced 
by Pati and Salam. At the intersection of these two defects a point 
defect arises with the 3-2-1 symmetry of the standard model, together 
with an extra $U(1)_X$ symmetry.}
\label{fig:6D-SO10}
\end{center}
\end{figure}

In our scheme, the leading power correction to the gauge couplings, 
which is not a calculable quantity in the effective field theory, 
is universal and thus does not contribute to the low energy 
prediction~\cite{Hall:2001pg, Nomura:2001mf}. The relative running 
of the gauge couplings, which is crucial for the prediction, is then 
reliably computed if the volume of the defects is sufficiently small 
-- that is, if the defects can effectively be viewed as points in 
the bulk: $d' = 0$. In such a setup, the low energy QCD coupling can 
be predicted in terms of the geometry of the bulk and the boundary 
conditions imposed on the bulk fields $\phi(y)$:
\begin{equation}
  \alpha_s = \alpha_s \left( d,\, {\rm geometry},\, 
  {\rm boundary \,\, conditions},\, \phi(y) \right).
\label{eq:genpred}
\end{equation}
As we go to a higher dimensional bulk, many more possibilities
open up for the structure of gauge symmetry breaking by boundary
conditions. A variety of defects can be incorporated. An example of a
$G = SO(10)$ theory in a 2D bulk~\cite{Hall:2001xr} is shown 
in Figure~\ref{fig:6D-SO10}.

There are two 1D defects: a line where the gauge symmetry is 
$SU(5) \times U(1)_X$ and a line where the gauge symmetry is the
Pati-Salam subgroup $SU(4)_C \times SU(2)_L \times SU(2)_R$. The
intersections of these lines gives a point defect where the 
reduced symmetry is that of the standard model, augmented by 
an extra $U(1)_X$. Other $SO(10)$ models with a 2D bulk are also 
possible~\cite{Asaka:2001eh, Hall:2001xr}, but generically it is 
hard, though not impossible~\cite{Hebecker:2001jb}, to reduce the 
rank by boundary conditions.

The prediction from gauge coupling unification differs in all these
variations. As shown in section~\ref{sec:minmodel}, the minimal $SU(5)$ 
theory with a single extra dimension agrees most precisely with data, 
and this suggests that in theories with higher gauge unification, such 
as $SO(10)$ in 6D, one dimension of the box is larger than the 
others~\cite{Hall:2002qw}.

\subsection{Split multiplets}

If the unified multiplet $H$ which contains the Higgs doublet of the
standard model is described by a bulk field $H(y)$, then the ``machine''
of Figure~\ref{fig:machine} will automatically lead to a splitting of 
order $1/R$ between the $SU(2)$ doublet component, $h_2$, and other
components. The ``doublet-triplet splitting'' puzzle of 4D unified 
theories is gone: indeed, mass splittings between components of bulk 
multiplets are unavoidable!  This phenomenon has been known 
since the mid 80s~\cite{Candelas:1985en, Dixon:jw}. However, the 
implementation of this for the Higgs multiplet is not obvious --- 
there are many possible geometries and boundary conditions; and with 
supersymmetry in the higher dimensional bulk, as needed for gauge 
coupling unification, the field $H(y)$ contains many more superpartners 
than the 4D case. In 2000, Kawamura discovered an extremely elegant 
solution~\cite{Kawamura:2001ev}: he studied a $G = SU(5)$ theory 
in 5D and constructed boundary conditions which broke the gauge 
symmetry to 3-2-1 such that only the massless modes of $H(y)$ were 
the Higgs doublets and their usual 4D superpartners. The colored 
triplet components, whose exchange leads to proton decay, and all 
the 5D superpartners were found to only have massive modes.

\begin{figure}
\begin{center} 
\begin{picture}(150,40)(10,-20)
  \Line(0,0)(150,0) \Text(75,-7)[t]{\large $SU(5)$}
  \Vertex(0,0){2}   \Text(0,8)[br]{$SU(5)$}
  \Vertex(150,0){2} \Text(150,8)[bl]{3-2-1}
\end{picture}
\caption{The minimal structure: $SU(5)$ gauge symmetry in a 1D bulk
with a single 3-2-1 point defect.}
\label{fig:5D-SU5}
\end{center}
\end{figure}
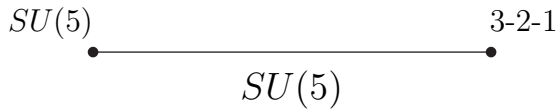

In our language of defects the geometry following from his boundary 
conditions appears almost trivially simple: a 1D $SU(5)$ bulk, having 
a 3-2-1 defect at one boundary but not at the other, as shown in 
Figure~\ref{fig:5D-SU5}. This case was also illustrated in 
Figures~\ref{fig:picture} and \ref{fig:orbifold}.

One might object that the whole idea of geometrical defects in gauge
transformations is ugly, destroying the beauty of complete symmetry 
in the underlying theory broken only spontaneously by a dynamical
choice of the vacuum. It is our contention that nature appears to
prefer such defects, and the first hint of this was the understanding
of the light Higgs doublets given by Kawamura.

\subsection{Proton decay}

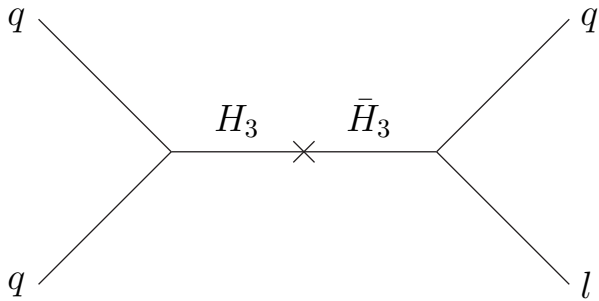
\begin{figure}
\begin{center} 
\begin{picture}(200,100)(0,0)
  \Line(0,0)(50,50)      \Text(-5,0)[r]{\large $q$}
  \Line(0,100)(50,50)    \Text(-5,100)[r]{\large $q$}
  \Line(200,0)(150,50)   \Text(205,0)[l]{\large $l$}
  \Line(200,100)(150,50) \Text(205,100)[l]{\large $q$}
  \Line(50,50)(150,50) \Line(96,46)(104,54) \Line(96,54)(104,46)
  \Text(75,58)[b]{\large $H_3$} \Text(125,58)[b]{\large $\bar{H}_3$}
\end{picture}
\caption{Baryon number violation is generated in 4D supersymmetric 
 unified theories via the exchange of massive colored Higgs triplets.}
\label{fig:proton-d}
\end{center}
\end{figure}

In 4D supersymmetric $SU(5)$ grand unified theories, the two Higgs
doublets, $h_2, \bar{h}_2$ are accompanied by their $SU(5)$ partners 
$H_3, \bar{H}_3$, which are color triplets. The exchange of the heavy
colored Higgs fermions yields a proton decay amplitude at dimension 
five via the diagram shown in Figure~\ref{fig:proton-d}. The cross on 
the internal line represents the Dirac mass that couples the fermions 
in $H_3$ and $\bar{H}_3$. In the minimal theory this is the only 
way these fermions can get heavy, and the model is excluded by the 
resulting large amplitude for proton decay~\cite{Goto:1998qg}.

What happens in higher dimensional unified theories with symmetry
breaking defects?  At first sight the situation looks very bad: 
although we understand why the Higgs triplets are heavy and the Higgs 
doublets are light, the mass of the Higgs triplets will be determined 
by geometry and will be of the order of the compactification scale, 
$1/R$. From gauge coupling unification we do not expect this to be 
large enough to avoid disastrous proton decay from $H_3$ fermion 
exchange. The origin of the masses of the modes in the KK towers is 
easily understood by considering the wavefunctions of the particles 
in a box with appropriate boundary conditions. For example, for the 
5D $SU(5)$ theory, the curves of Figure~\ref{fig:orbifold} can be 
reinterpreted, with the solid curves being the wavefunction of the 
Higgs doublet modes and the dotted curves being the wavefunctions of 
the Higgs triplet modes. Since $E^2 = p_x^2 + p_y^2$, the momentum 
$p_y$ is interpreted as a mass in 4D, so that the only massless mode 
is that of the Higgs doublet having the flat wavefunction. All the 
Higgs triplet modes have non-zero $p_y$ because boundary conditions 
force their wavefunctions to vanish at the 3-2-1 defect, and the 
corresponding 4D fields have masses of order $1/R$. In the minimal 
theory discussed in the next section $1/R \approx 10^{15}~{\rm GeV}$, 
which is less than the unified mass scale in the conventional 4D 
theory. Hence one might expect a very large proton decay amplitude.

In 5D, the form of the mass terms for the $H_3$ and $\bar{H}_3$ 
fermions is dictated by the higher dimensional spacetime symmetry of 
the bulk. Since the smallest fermion representation in 5D is a Dirac 
fermion, both $H_3$ and $\bar{H}_3$ fermions are accompanied by their 
conjugated fermions, $H^c_3$ and $\bar{H}^c_3$. The 5D kinetic terms 
for these fermions contain $H_3 \partial_y H^c_3$ and $\bar{H}_3 
\partial_y \bar{H}^c_3$. From the viewpoint of 4D the masses arise 
from $\partial_y$, so that the Dirac mass for $H_3$ couples it to 
$H^c_3$ rather than to $\bar{H}_3$. The cross in the diagram of 
Figure~\ref{fig:proton-d} does not exist; $\bar{H}_3$ must be replaced
by $H^c_3$. On the other hand, an $R$ symmetry arising from higher 
dimensional supersymmetry forbids any coupling of $H^c_3$ to quarks 
and leptons. Hence, we find that the proton decay amplitude from the 
exchange of the color triplet Higgs fermions necessarily vanishes in 
higher dimensional unified theories~\cite{Hall:2001pg}.

We have shown that the absence of all proton decay from operators in the 
low energy theory of dimension four or five is guaranteed by an $R$ 
symmetry~\cite{Hall:2001pg, Hall:2001xb}. Hence the leading contribution 
will come at dimension six from the exchange of the heavy $X$ gauge 
bosons. This depends sensitively on the mass of these gauge bosons, 
which is also given by $1/R$. The precise value of $1/R$ is model 
dependent, and we will return to this issue in section~\ref{sec:minmodel}.

\subsection{Quark-lepton mass relations}

So far we have assumed that particles are free to propagate
throughout the volume of the bulk. However, it may be that some
particles are restricted to subspaces of the bulk. For example, in the
box of Figure~\ref{fig:extra-3D}, quarks and leptons could be chosen 
to propagate in the entire 3D bulk, on a given 2D surface, a 1D line, 
or they may even be restricted to a point. A quark or lepton which 
propagates on a defect with lower gauge symmetry will only feel this 
lower symmetry: it will not live in a multiplet of the higher gauge 
symmetry of the full bulk. It would therefore seem less attractive 
to place quarks and leptons precisely on a 3-2-1 defect, since one 
would lose the immediate understanding of the gauge quantum numbers 
of a generation given by the higher gauge symmetry.\footnote{
On the other hand, placing the Higgs on a 3-2-1 defect is an 
alternative way to understand the absence of a color triplet Higgs 
in the low energy theory~\cite{Hebecker:2001wq}.}

What distinguishes one generation from another? Could it be that 
they propagate in differing numbers of dimensions in the bulk?  This
is an attractive idea because it leads to a geometrical understanding
of the hierarchy between the masses of the generations. The quark and
lepton masses arise from Yukawa couplings, but these interactions are
forbidden by supersymmetry in dimensions higher than 4. Hence the
Yukawa coupling between fermion $\psi_i$ and fermion $\psi_j$ 
must be located at a point $y_0$ on the surface of the box where 
the higher dimensional Lorentz and supersymmetry is broken: 
${\cal L}_{\rm Yukawa} = \delta^d(y - y_0) \psi_i \psi_j H(y)$. 
Since the quarks and leptons that we observe are zero mass modes in 
the box, they have flat wavefunctions with normalizations $1/\sqrt{V}$, 
where $V$ is the volume of the subspace of the bulk in which they 
propagate. Integrating over the volume of the bulk to get the 
equivalent 4D theory then leads to an entry in the fermion mass matrix
\begin{equation}
  m_{ij} \propto {1 \over \sqrt{V_i V_j}}, 
\label{eq:fermmasshier}
\end{equation}
providing both subspaces cover the point $y_0$. 

The above relation implies that the heaviest fermions propagate in only 
a small subspace of the bulk. This makes it likely that their Yukawa 
coupling is located far from the $G$-breaking defects, so that they will 
exhibit unified mass relations between quarks and leptons. On the other 
hand, lighter fermions must live in a larger subspace of the bulk, and 
are more likely to propagate past the defects. Yukawa couplings located 
on these defects will destroy any unified mass relations. {\it Unified 
mass relations between quarks and leptons are expected only for the 
heaviest generation}~\cite{Hall:2001xb, Hall:2001zb, Hall:2001rz, 
Nomura:2001tn}.

To conclude: higher dimensional grand unified theories with symmetry
breaking defects offer a remarkable possibility. The conventional
successes of grand unified theories (quark-lepton gauge quantum
numbers, gauge coupling unification, and mass relations for heavy
fermions) can be retained as phenomena of the symmetrical bulk, while
conventional difficulties (mass splitting between $h_2$ and $H_3$,
proton decay, and light fermion mass relations) are automatically
resolved as phenomena of the defects.

\section{The Minimal Model}
\label{sec:minmodel}

Up to now we have concentrated on the conceptual advantages of higher 
dimensional unified theories.  We now show that these theories are
remarkably predictive if they are valid over a large energy range,
{\it i.e.} if $M_s/M_c$ is large, and present a minimal 
model~\cite{Hall:2001xb} which is highly successful in describing 
physics over a wide energy interval between $M_s$ and $M_c$.

\subsection{Preferred by gauge coupling unification}

{\it The} numerical test for any unified theory is gauge coupling
unification, so we intend to use this as a tool to guide us in
searching for a particular higher dimensional geometry. Recall from
Figure~\ref{fig:alphas_1}: while conventional supersymmetric 
unification does well, it is not perfect. We have also seen from 
Figure~\ref{fig:machine} that the box and boundary conditions are 
a ``machine'' for creating KK towers of particles. Could the 
difference between the central value of the conventional prediction, 
$\alpha_s(M_Z) = 0.130$, and experiment, $\alpha_s(M_Z) = 0.117$, be 
due to the virtual effects of these KK modes? If so, are there any 
geometries that are simple enough that they are numerically predictive?

We have performed a detailed study of supersymmetric theories with 
$d$ extra dimensions with equal radii. For $d \geq 3$ there are no
corrections from the KK modes because of cancellations forced by the
large amount of supersymmetry in higher dimensions. For $d=1,2$ the
leading logarithmic correction is~\cite{Hall:2001xb}
\begin{equation}
  \delta\alpha_s \simeq -\frac{6}{7n \pi} \alpha_s^2 \ln\frac{M_s}{M_c},
\label{eq:as-formula}
\end{equation}
where $n=2$ for $d=1$, and  for $d=2$ it is a positive integer $\geq 2$,
describing the geometry of the box. Here, $\delta\alpha_s$ is defined 
by the difference of our prediction, $\alpha_s^{\rm KK}$, and the 
conventional prediction, $\alpha_s^{\rm SGUT}$: $\delta\alpha_s = 
\alpha_s^{\rm KK} - \alpha_s^{\rm SGUT}$. This result applies only 
if the Higgs doublets propagate in the bulk, and are not contained in 
the vector multiplet; otherwise the sign of the correction is changed,
increasing the discrepancy with data. While this result is very 
simple, and the sign is very encouraging, apparently we cannot 
evaluate it numerically because of the unknowns $M_s/M_c$ and $n$. 
However, using our assumption that the theory is strongly coupled 
at the fundamental scale, $M_s/M_c$ can be estimated as in 
Eq.~(\ref{eq:strongcoup}). From this we discover that for most 
values of $(d,n)$ the correction $|\delta\alpha_s|$ is too small to 
give perfect agreement with data. Only in the case that it is maximized 
does the central value of the theoretical prediction agree with data, 
and this occurs for the simplest case of a single extra dimension, 
$d=1$ (hence $n=2$). In this case the unified gauge group should 
be $SU(5)$, since larger unified groups cannot be broken by boundary 
conditions in a single extra dimension to 3-2-1-$G'$, so that gauge 
coupling unification would depend on further symmetry breaking and 
predictivity would be lost. These considerations lead to the effective 
theory below $M_s$ as given in Figure~\ref{fig:eff-theory}.
\begin{figure}
\begin{center} 
\begin{picture}(275,200)(-40,0)
  \LongArrow(0,0)(0,200) \Text(-5,200)[r]{energy}
  \DashLine(0,80)(200,80){4} \DashLine(0,160)(200,160){4}
  \Text(-8,80)[r]{$M_c \simeq 5 \times 10^{14}~{\rm GeV}$}
  \Text(-8,160)[r]{$M_s \simeq 1 \times 10^{17}~{\rm GeV}$}
  \Text(100,180)[]{\large UV (string?) theory}
  \Line(50,120)(150,120) \Text(100,117)[t]{\large $SU(5)$}
  \Vertex(50,120){2}  \Text(50,127)[br]{$SU(5)$}
  \Vertex(150,120){2} \Text(150,127)[bl]{3-2-1}
  \Text(100,40)[]{\large MSSM}
\end{picture}
\caption{The scheme preferred by gauge coupling unification: the 
 minimal supersymmetric standard model (MSSM) is the effective theory 
 up to $M_c \approx 5 \times 10^{14}~{\rm GeV}$, while the effective 
 theory for the next factor of $200$ in energy is the minimal 5D $SU(5)$ 
 theory with a single 3-2-1 defect.}
\label{fig:eff-theory}
\end{center}
\end{figure}
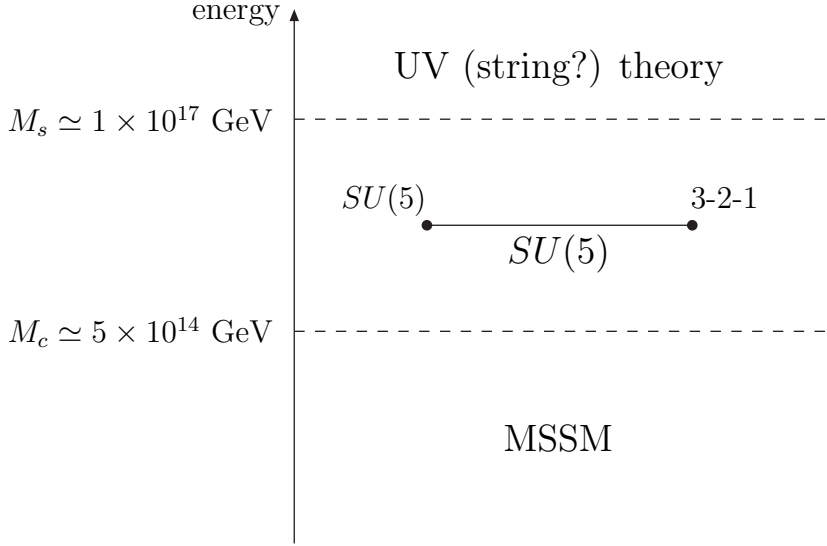
\begin{figure}
\begin{center} 
\begin{picture}(275,300)(-40,0)
  \LongArrow(0,0)(0,300) \Text(-5,300)[r]{$\alpha_s(M_Z)$}
  \Line(-3,15)(3,15)   \Text(-8,16)[r]{$0.060$}
  \Line(-3,45)(3,45)
  \Line(-3,75)(3,75)   \Text(-8,76)[r]{$0.080$}
  \Line(-3,105)(3,105)
  \Line(-3,135)(3,135) \Text(-8,136)[r]{$0.100$}
  \Line(-3,165)(3,165)
  \Line(-3,195)(3,195) \Text(-8,196)[r]{$0.120$}
  \Line(-3,225)(3,225)
  \Line(-3,255)(3,255) \Text(-8,256)[r]{$0.140$}
  \Line(-3,285)(3,285)
% experiment
  \Text(210,186)[l]{$\alpha_s^{\rm exp}$}
  \DashLine(0,180)(200,180){4}
  \DashLine(0,192)(200,192){4}
% GUT
  \Text(50,285)[b]{$\alpha_s^{\rm GUT}$}
  \DashLine(50,60)(50,72){1} \Vertex(50,66){3}
  \DashLine(45,60)(55,60){1} \DashLine(45,72)(55,72){1} 
% SUSY GUT
  \Text(100,285)[b]{$\alpha_s^{\rm SGUT}$}
  \DashLine(100,216)(100,234){1}
  \DashLine(95,216)(105,216){1} \DashLine(95,234)(105,234){1}
  \Line(100,213)(100,237) \Vertex(100,225){3}
  \Line(95,213)(105,213) \Line(95,237)(105,237)
% KK GUT
  \Text(150,285)[b]{$\alpha_s^{\rm KK}$}
  \DashLine(150,180)(150,198){1}
  \DashLine(145,180)(155,180){1} \DashLine(145,198)(155,198){1}
  \Line(150,177)(150,201) \Vertex(150,189){3}
  \Line(145,177)(155,177) \Line(145,201)(155,201)
  \LongArrow(53,80)(93,206) \Text(75,140)[tl]{SUSY log}
  \LongArrow(108,219)(142,195) \Text(121,215)[bl]{KK log}
\end{picture}
\caption{The predictions for $\alpha_s(M_Z)$ in the three frameworks: 
 non-supersymmetric grand unification $\alpha_s^{\rm GUT}$, 
 supersymmetric grand unification $\alpha_s^{\rm SGUT}$, and 
 higher dimensional grand unification $\alpha_s^{\rm KK}$. 
 Solid error bars represent the threshold corrections from the 
 superpartner spectrum. Dotted error bars for $\alpha_s^{\rm GUT}$ and 
 $\alpha_s^{\rm SGUT}$ represent threshold corrections from
 the unified scale corresponding to a heavy ${\bf 5} + \bar{\bf 5}$
 representation with unit logarithmic mass splitting between doublets
 and triplets. The dotted error bar for $\alpha_s^{\rm KK}$ is the
 theoretical uncertainty (other than from superpartner masses) for 
 our theory, as estimated in Ref.~\cite{Hall:2001xb}.}
\label{fig:alphas_2}
\end{center}
\end{figure}
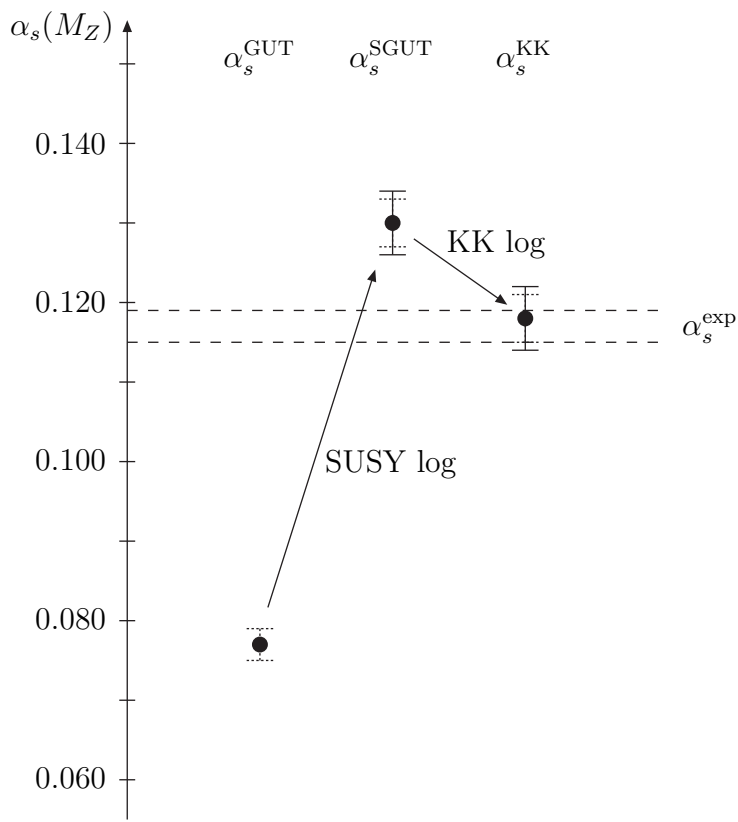
Note that we are able to go much further than conventional 
supersymmetric unification which simply identifies a single scale 
$M_u \simeq 2 \times 10^{16}~{\rm GeV}$ as the threshold for unified 
physics. We can determine both the compactification scale, 
$M_c \simeq 5 \times 10^{14}~{\rm GeV}$, and the scale of strong 
coupling, $M_s \simeq 1 \times 10^{17}~{\rm GeV}$, and consequently 
the masses of all the KK modes of gauge bosons, Higgs and matter in 
this energy interval.\footnote{
Here and below, $M_c$ represents the length scale of the extra 
dimension, $M_c = (\pi R)^{-1}$, which is denoted as $M'_c$ in 
Refs.~\cite{Hall:2001xb, Hall:2002ci}.}
Apart from discrete choices, such as the location of the quarks and 
leptons of the various generations, we determine the entire effective 
theory that is valid over an energy range spanning a factor of $200$. 

In Figure~\ref{fig:alphas_2} we show the effect of the logarithm from 
the KK modes in the minimal model --- the prediction is strikingly 
successful. Because the higher dimensional theory is valid over 
such a large energy interval, the uncertainties to this correction 
are small, as shown in the figure. The dominant uncertainty in the 
prediction now comes from the supersymmetric threshold, which 
ultimately will be fixed by data.

At energies approaching $M_s$, our 5D effective theory will break
down. It could be that a higher dimensional structure emerges, such 
as that of Figure~\ref{fig:6D-SO10}, with the vertical dimension much 
less than the horizontal one. At this scale it may be possible to 
interpret the Higgs as arising from a vector multiplet, for example 
as a component of the higher dimensional gauge field~\cite{Hall:2002qw, 
Hall:2001zb}.

\subsection{Yukawa coupling unification}

In section~\ref{sec:newphysics} we have argued that heavier quarks 
and leptons should propagate in subspaces of the bulk with lower 
dimension.  In the minimal theory the heavy third generation should 
reside at a boundary of the fifth dimension. To retain the $SU(5)$ 
understanding of quantum numbers, this should be the ``$SU(5)$ 
boundary'' rather than the ``3-2-1 boundary'', giving the usual 
tree-level $SU(5)$ mass relation: $m_b = m_\tau$. In conventional 
unified theories the corrections to this relation from running of 
the Yukawa couplings can be accurately computed, but there are also 
corrections from both unknown physics at the unified scale, and from 
supersymmetric corrections at the weak scale. In our theory the 
unified physics is known, and hence we can compute the corrections
at the unified scale. The resulting correction to the prediction 
for the $b$ quark mass is~\cite{Hall:2002ci}
\begin{equation}
  \frac{\delta m_b}{m_b} 
    \simeq - \frac{5(4g^2 - y_t^2)}{112\pi^2} \ln\frac{M_s}{M_c}.
\label{eq:delta-mb-Mc}
\end{equation}
As in Eq.~(\ref{eq:as-formula}) for the radiative corrections to 
$\alpha_s$, the sign improves agreement with data. However, unlike 
gauge coupling unification, the prediction for the $b$ quark mass 
receives large supersymmetric corrections~\cite{Hall:1993gn}, and 
hence we leave the comparison with data until section~\ref{sec:expsig} 
where we incorporate supersymmetry breaking.

Since the minimal theory has only a single extra dimension, all of
flavor cannot be understood in terms of volume factors. Nevertheless,
some of the quarks and leptons will have suppressed masses because
they propagate in the bulk, and these light fermions will not exhibit
unified mass relations. For example, if the right-handed up quark, $u$, 
and electron, $e$, reside in the bulk they will be the zero modes of 
a ${\bf 10}$-dimensional representation of $SU(5)$: $T(u,e)$, where 
the zero modes are written in parenthesis. However, in this case 
a light quark doublet $q$ does not arise from this multiplet; rather 
it arises from another ${\bf 10}$-plet having boundary conditions 
with opposite sign: $T'(q)$. This may appear to be a step backwards 
--- the light quarks and leptons are not as unified as in conventional 
$SU(5)$. However, it can also be viewed as a virtue: disastrous unified 
mass relations for light matter are avoided, while the understanding 
for quark and lepton quantum numbers is preserved. To see the absence 
of a mass relation, consider the fermion Yukawa couplings in the case 
where the right-handed down quark, $d$, and the lepton doublet, $l$, 
are unified into the same ${\bf 5}$-dimensional multiplet: $F(d,l)$.
In this case the mass terms for the electron and down quark arise from 
two distinct interactions $TF\bar{H}$ and $T'F\bar{H}$, respectively, 
so there is clearly no $SU(5)$ relation between these masses.

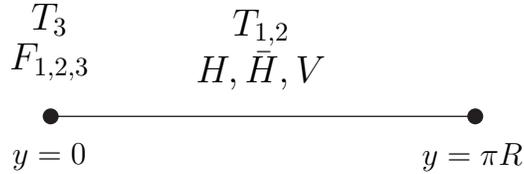
\begin{figure}
\begin{center} 
\begin{picture}(100,70)(150,20)
  \Line(120,50)(280,50)
  \Vertex(120,50){3} \Text(120,40)[t]{$y=0$}
  \Vertex(280,50){3} \Text(280,40)[t]{$y=\pi R$}
  \Text(120,83)[b]{\large $T_3$} \Text(120,68)[b]{\large $F_{1,2,3}$}
  \Text(200,81)[b]{\large $T_{1,2}$} 
  \Text(200,63)[b]{\large $H, \bar{H}, V$}
\end{picture}
\caption{Locations of $SU(5)$ matter, Higgs and gauge multiplets 
 in the fifth dimension.}
\label{fig:theory}
\end{center}
\end{figure}
One plausible possibility is that the lightest two generations 
of ${\bf 5}$-plets $F_{1,2}$ are located on the boundary, while 
the corresponding ${\bf 10}$-plets $T_{1,2}$ are bulk 
modes~\cite{Hall:2002ci}, as depicted in Figure~\ref{fig:theory}. 
There is no flavor distinction between the three $F$, so that large 
neutrino mixing angles are expected. Small neutrino masses can be 
understood by introducing right-handed neutrinos $N_{1,2,3}$, through 
the conventional see-saw mechanism~\cite{Seesaw}. Flavor hierarchies 
in masses and mixings arise from the $T$, both from volume 
factors~\cite{Hall:2001xb, Hall:2001zb, Hall:2001rz, Nomura:2001tn} 
and from distortions of the wavefunctions caused by bulk mass 
terms~\cite{Hebecker:2002re}. Potential brane-localized anomalies 
are canceled by a certain bulk term made out of the bulk gauge 
field~\cite{Arkani-Hamed:2001is}. This gives a larger hierarchy for 
up-type quark masses (from $TT$) than for down-type and charged 
lepton masses (from $TF$). The absence of unified relations amongst 
the lighter two generation quarks and leptons is entirely due to 
$T_{1,2}$ (which really represent $T_{1,2}(u,e)$ and $T'_{1,2}(q)$).

\subsection{Predictions for proton decay}

In section~\ref{sec:newphysics} we saw that higher dimensional theories 
possess a $U(1)_R$ symmetry that forbids proton decay from operators of 
dimension five resulting from the exchange of the colored triplet Higgs.
This $R$ symmetry arises from a phase rotation of the coordinates of
superspace --- it is an unavoidable consequence of supersymmetry in the
higher dimensional bulk. In 4D supersymmetric theories one must impose
a discrete symmetry by hand to avoid baryon- and lepton-number 
violation at the weak scale arising from the superpotential 
interactions $udd, qdl, lle$ and $lh$. In higher dimensional theories 
this parity can be understood as a subgroup of the continuous $R$ 
symmetry. Of course, one cannot prove that these interactions must be 
absent --- after all there might be an $R$ symmetry breaking defect 
on the boundary; however, we can say that in higher dimensional 
theories there is a very plausible origin for the conventional 
discrete symmetry.

Do these theories predict proton stability? No --- in general the heavy 
gauge bosons of the unified theory can induce proton decay, and hence we
must study the masses and couplings of these gauge bosons in models of
interest. In the 5D $SU(5)$ theory there is a KK tower of $X$ gauge
bosons with the lightest mode having mass $1/2R = \pi M_c/2$. We have 
seen that a calculable weak mixing angle requires a large value of 
$M_s/M_c$, and therefore a small value for $M_c$. The observed values 
of the gauge couplings strongly suggest that $M_s$ is the scale of 
strong coupling, so that $M_c \simeq 5 \times 10^{14}~{\rm GeV}$. 
Hence this gauge boson has about the same mass as in the 4D 
Georgi-Glashow $SU(5)$ theory. It appears that we have come around 
a full circle, and are excluded, like the non-supersymmetric $SU(5)$ 
theory, by searches for proton decay. However, we have not yet 
investigated the couplings of $X$ in the 5D theory. We have argued 
that since the electron and up quark are so light they should reside 
in the bulk; thus these states are described by two ${\bf 10}$-plets 
$T_1(u,e)$ and $T'_1(q)$. This means that the conventional interactions 
of the $X$ boson, $q^\dagger u X, e^\dagger q X$ are not generated by 
the bulk gauge interactions of the 5D theory, at least for the lightest 
generation, and hence in the absence of CKM mixing between the 
generations the proton would be stable~\cite{Hall:2001pg}. The mode 
expected from CKM mixing is $p \rightarrow K^+ \bar{\nu}$, but 
the rate is now highly dependent on the flavor structure of the 
theory~\cite{Nomura:2001tn, Hebecker:2002rc}, and while the rate 
is no longer too large, it is not guaranteed to be in reach of 
future detectors.

Remarkably there is an additional source for the $q^\dagger u X,
e^\dagger q X$ interactions. They result from a boundary localized 
contribution to the gauge interactions, and therefore have a size
which is suppressed relative to the usual gauge coupling by the volume 
factor $M_c/M_s$. The proton lifetime cannot be precisely predicted 
since the boundary gauge interaction involves a dimensionless
coupling that is not predicted. If this coupling is of order unity
then $\tau_p \approx 10^{34}~{\rm years}$, with comparable branching 
ratios for the decay modes $e^+ \pi^0, \mu^+ \pi^0, e^+ K^0, \mu^+ K^0, 
\pi^+ \bar{\nu}$ and $K^+ \bar{\nu}$~\cite{Hall:2002ci}. The most 
promising discovery mode is $e^+\pi^0$. A large mixing angle between 
$F_1$ and $F_2$ implies that the $e^+ \pi^0$ and $\mu^+ \pi^0$ modes 
have comparable branching ratios and that 
\begin{equation}
  \frac{\Gamma(p \rightarrow \mu^+\pi^0)}{\Gamma(p \rightarrow e^+\pi^0)} 
\simeq 
  \frac{\Gamma(p \rightarrow e^+K^0)}{\Gamma(p \rightarrow \mu^+K^0)},
\label{eq:pratios}
\end{equation}
independent of the sizes of hadronic matrix elements~\cite{Hall:2002ci}.
This analysis for proton decay depends on matter location, but is 
completely independent of supersymmetry breaking.

\section{Supersymmetry Breaking}
\label{sec:expsig}

Theories with weak scale supersymmetry will lead to a plethora of 
new particles and couplings to be measured in the TeV domain. The
supersymmetry breaking interactions, like the gauge and Yukawa
interactions, may encode information about the unified 
theory~\cite{Hall:1985dx}. For this to happen they must remain as 
local interactions up to the unification scale --- we say that the 
messenger scale of supersymmetry breaking must be at least as large 
as the unification scale. If this happens in higher dimensional 
theories, then the soft supersymmetry breaking operators will provide 
a window on the physics of the bulk~\cite{Hall:2002ci}. In particular, 
the flavor symmetry of the bulk $SU(5)$ gauge interactions 
$U(3)_T \times U(3)_F$ will be modified by the locations of the 
three $T$ and three $F$ fields. This will lead to non-universal 
squark and slepton masses, and to flavor changing neutral currents
from superpartner exchange, allowing experiment to probe the geometry
and matter locations in the bulk.

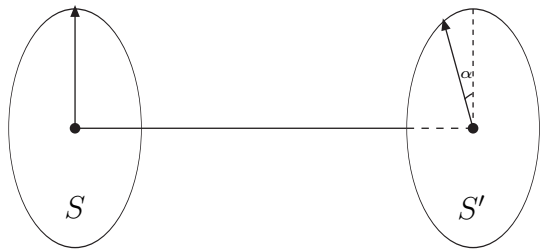
\begin{figure}
\begin{center} 
\begin{picture}(150,90)(10,-45)
  \Line(0,0)(125,0) \DashLine(125,0)(150,0){3}
  \Vertex(0,0){2}   \Oval(0,0)(45,25)(0)   \LongArrow(0,0)(0,45)
  \CArc(154,4)(10,115,136)  \DashLine(150,0)(150,45){2}
  \Vertex(150,0){2} \Oval(150,0)(45,25)(0) \LongArrow(150,0)(139,40)
  \Text(148,19)[b]{\tiny $\alpha$}
  \Text(0,-30)[]{$S$} \Text(150,-30)[]{$S'$} 
\end{picture}
\caption{The breaking of supersymmetry by the misalignment of
 boundary defects.}
\label{fig:SUSY-Br}
\end{center}
\end{figure}
Such signals will depend on the mechanism for supersymmetry 
breaking. In this section we discuss the possibility that boundary 
conditions in the bulk break supersymmetry as well as the unified 
gauge symmetry~\cite{Barbieri:2001yz}. In the minimal 5D $SU(5)$ model 
there is a unique way to accomplish such a breaking~\cite{Barbieri:2001dm}. 
In the 5D bulk there are two independent supersymmetry transformations
and they form a doublet of the $SU(2)$ $R$ symmetry. However, the
boundaries are four dimensional and can support only a single
supersymmetry. Thus the boundaries can be viewed as defects in the space 
of supersymmetry transformations, and the key question is whether the 
two supersymmetries, $S$ and $S'$, respected by the two boundaries 
are the same or not, as illustrated in Figure~\ref{fig:SUSY-Br}. 
In the limit that the relative orientation angle $\alpha$ vanishes, 
$S' = S$ and the single supersymmetry $S$ is preserved everywhere and 
must therefore be a symmetry of the low energy 4D theory. On the other 
hand, if there is a misalignment of the two defects characterized by 
a non-zero value of $\alpha$, then supersymmetry is broken in the low 
energy theory, with the supersymmetry breaking mass scale given by 
$\tilde{m} = \alpha M_c$. The appearance of the continuous parameter 
$\alpha$ describing the defects represents an important difference 
compared with the case of $SU(5)$ gauge symmetry breaking, where 
the breaking scale is necessarily $M_c$. The parameter $\alpha$ can 
also be viewed as the vacuum condensation of a component of the higher 
dimensional gravitational supermultiplet~\cite{Marti:2001iw}, implying 
that the supersymmetry breaking is spontaneous and the hierarchy of 
the scales, $\tilde{m} \ll M_c$, can naturally be obtained.

The tree-level form of the supersymmetry breaking soft operators is 
very simple~\cite{Hall:2002ci}. All bulk superpartners have mass 
$\tilde{m}$, all boundary-located superpartners are massless, and the 
size of the trilinear scalar interaction is $\tilde{m}, 2\tilde{m}$ or 
$3\tilde{m}$, counting the number of bulk scalars:
\begin{eqnarray}
  {\cal L}_{\rm soft} &=& 
    - \frac{1}{2} \left( \tilde{m} \lambda \lambda + {\rm h.c.} \right)
    - \tilde{m}^2 h^\dagger h 
    - \tilde{m}^2 \tilde{f}_B^\dagger \tilde{f}_B 
\nonumber\\
 && + \left( y_f \tilde{m} \tilde{f}_b \tilde{f}_b h 
        + 2 y_f \tilde{m} \tilde{f}_B \tilde{f}_b h
        + 3 y_f \tilde{m} \tilde{f}_B \tilde{f}_B h + {\rm h.c.} \right),
\label{eq:soft}
\end{eqnarray}
where $\lambda$, $h$, $\tilde{f}_B$ and $\tilde{f}_b$ collectively 
represent the gauginos, two Higgs doublets, squarks/sleptons in 
the bulk and squarks/sleptons on the brane, respectively, and $y_f$ 
is the value of the corresponding Yukawa coupling. Since supersymmetry 
breaking effects from boundary conditions are shut off above the 
compactification scale, the soft supersymmetry breaking masses in 
Eq.~(\ref{eq:soft}) must be regarded as the running mass parameters 
at the compactification scale $M_c$. Incidentally, this type of 
theory can also generate the weak-scale $\mu$ and $B$ terms naturally 
through the vacuum readjustment mechanism~\cite{Hall:2002up}.

Applying these results to the minimal model yields two immediate 
consequences: first the flavor changing neutral current effects induced 
by superpartner exchange are too large unless the squarks (sleptons) of 
the first two generations are degenerate. This means that $T_1$ and 
$T_2$ must have the same location, and similarly for $F_1$ and $F_2$. 
When coupled with constraints from proton decay and from fermion mass 
relations, the location assignments of Figure~\ref{fig:theory} become 
unique. Hence, we are now able to predict that the neutrino mixing 
angles are large. Secondly the Higgs potential and superpartner masses 
are determined in terms of only three parameters $\tilde{m}$, $\mu$ and 
$B$. After fixing the $Z$ mass the two independent parameters can be 
taken to be $\tilde{m}$ and $\tan\beta$, the ratio of the two Higgs 
vacuum expectation values. The entire superpartner and Higgs spectrum 
is shown in Table~\ref{table:spectrum} for two representative values 
of these parameters.  The mass of the lightest Higgs boson, $h$, 
includes one-loop radiative corrections from top quarks and squarks.
The sign and large size of the relevant scalar trilinear of 
Eq.~(\ref{eq:soft}) makes these corrections large and the resulting 
Higgs boson becomes relatively heavy. The squarks and sleptons from 
$T_3$ and $F_{1,2,3}$ are relatively light as they reside on the 
boundary and do not get mass at tree level. For example, this makes 
$\tilde{l}$ lighter than $\tilde{e}$. The two lightest superpartners 
are the scalar tau and the bino. Either could be the LSP, and hence 
collider signals involve either stable charged particle tracks or 
missing transverse energy~\cite{Hall:2002ci}.
\begin{table}
\begin{center}
\begin{tabular}{|c|c|c|}  \hline 
 $\tilde{m}$ & $300$ & $400$ \\
 $\tan\beta$ & $5$   & $10$  
\\ \hline
 $\tilde{g}$            & $699$ & $911$ \\
 $\tilde{\chi}^{\pm}_1$ & $251$ & $334$ \\
 $\tilde{\chi}^{\pm}_2$ & $427$ & $531$ \\
 $\tilde{\chi}^0_1$     & $130$ & $175$ \\
 $\tilde{\chi}^0_2$     & $251$ & $334$ \\
 $\tilde{\chi}^0_3$     & $417$ & $518$ \\
 $\tilde{\chi}^0_4$     & $422$ & $528$ 
\\ \hline
 $\tilde{q}$ & $701$ & $915$ \\
 $\tilde{u}$ & $675$ & $880$ \\
 $\tilde{d}$ & $602$ & $780$ \\
 $\tilde{l}$ & $209$ & $277$ \\
 $\tilde{e}$ & $317$ & $422$ 
\\ \hline
 $\tilde{t}_1$    & $425$ & $547$ \\
 $\tilde{t}_2$    & $619$ & $780$ \\
 $\tilde{b}_1$    & $563$ & $727$ \\
 $\tilde{b}_2$    & $601$ & $774$ \\
 $\tilde{\tau}_1$ & $106$ & $126$ \\
 $\tilde{\tau}_2$ & $214$ & $280$
\\ \hline
 $h$       & $118$ & $128$ \\
 $A$       & $552$ & $690$ \\
 $H^0$     & $553$ & $690$ \\
 $H^{\pm}$ & $558$ & $695$ 
\\ \hline
 $\alpha_s(M_Z)$ \{$\pm 0.003$\} & $0.119$ & $0.118$ \\
 $m_b(M_Z)$ \{$\pm 0.10$\}       & $3.37$  & $3.26$  
\\ \hline
 ${\rm Br}(\mu \rightarrow e\gamma)$                 & 
   $6\times10^{-12}$ & $8\times10^{-12}$ \\
 ${\rm Br}(\mu \rightarrow 3e)$                      & 
   $4\times10^{-14}$ & $5\times10^{-14}$ \\
 ${\rm Cr}(\mu \rightarrow e; {}^{48}_{22}{\rm Ti})$ & 
   $4\times10^{-14}$ & $5\times10^{-14}$ \\
 ${\rm Br}(\tau \rightarrow \mu\gamma)$              & 
   $1\times10^{-8}$  & $1\times10^{-8}$  
\\ \hline
\end{tabular}
\end{center}
\caption{Predictions for the superpartner spectrum, the Higgs spectrum, 
 gauge and Yukawa unification, and lepton flavor violating processes. 
 The predictions are for two representative values of $\tilde{m}$ and 
 $\tan\beta$, and all masses are given in GeV. Mass eigenvalues are 
 given for the gluino, $\tilde{g}$, the charginos, $\tilde{\chi}^\pm$, 
 the neutralinos, $\tilde{\chi}^0$, the squarks and sleptons of the 
 third generation, $\tilde{t}_{1,2}, \tilde{b}_{1,2}$ and 
 $\tilde{\tau}_{1,2}$, and the Higgs bosons, $h, A, H^0$ and $H^{\pm}$. 
 For the first two generations of squarks and sleptons the masses are 
 shown for $\tilde{q}, \tilde{u}, \tilde{d}, \tilde{l}$ and $\tilde{e}$ 
 and do not include contributions from electroweak $D$ terms.}
\label{table:spectrum}
\end{table}
\begin{figure}
\begin{center} 
\begin{picture}(250,245)(0,-100)
  \LongArrow(0,-100)(0,145) \Text(-5,145)[r]{$m_b(M_Z)$}
  \Line(-2,-90)(2,-90)
  \Line(-4,-75)(4,-75) \Text(-9,-74)[r]{$2.5$}
  \Line(-2,-60)(2,-60) \Line(-2,-45)(2,-45)
  \Line(-2,-30)(2,-30) \Line(-2,-15)(2,-15)
  \Line(-4,0)(4,0)     \Text(-9,1)[r]{$3.0$}
  \Line(-2,15)(2,15)   \Line(-2,30)(2,30)
  \Line(-2,45)(2,45)   \Line(-2,60)(2,60)
  \Line(-4,75)(4,75)   \Text(-9,76)[r]{$3.5$}
  \Line(-2,90)(2,90)   \Line(-2,105)(2,105)
  \Line(-2,120)(2,120) \Line(-2,135)(2,135)
  \LongArrow(0,0)(250,0) \Text(250,-2)[t]{$\tan\beta$}
  \Line(50,-2)(50,2)   \Text(51,-5)[t]{$5$}
  \Line(100,-2)(100,2) \Text(101,-5)[t]{$10$}
  \Line(150,-2)(150,2) \Text(151,-5)[t]{$15$}
  \Line(200,-2)(200,2) \Text(201,-5)[t]{$20$}
% experiment
  \Line(10,-45)(10,45) \Vertex(10,0){3}
  \Line(5,-45)(15,-45) \Line(5,45)(15,45)
  \DashLine(20,-45)(250,-45){4}
  \DashLine(20,45)(250,45){4}
  \Text(15,-30)[l]{exp}
% SUSY GUT
  \GOval(125,100)(12,75)(0){0.92}
  \Text(125,100)[]{SGUT}
% KK GUT
  \Line(50,40)(50,70) \Vertex(50,55){3}
  \Line(45,40)(55,40) \Line(45,70)(55,70)
  \Line(50,55)(200,6) 
  \DashLine(50,70)(200,21){2} \DashLine(50,40)(200,-9){2}
  \Line(200,-9)(200,21) \Vertex(200,6){3}
  \Line(195,-9)(205,-9) \Line(195,21)(205,21)
  \Text(210,18)[bl]{KK}
\end{picture}
\caption{A comparison of the prediction of the $b$ quark mass from 
 Yukawa unification with experiment. The prediction from conventional 
 4D unification, without supersymmetric threshold corrections is too 
 large. In the minimal 5D $SU(5)$ theory, the unified corrections 
 of Eq.~(\ref{eq:delta-mb-Mc}) bring the prediction within 1$\sigma$ 
 of the data. If supersymmetry is broken by boundary conditions, the 
 superpartner corrections are linear in $\tan\beta$ and bring complete 
 agreement with data. Values of $\tan\beta$ larger than about 20 are 
 disfavored by lepton flavor changing processes.}
\label{fig:b-tau}
\end{center}
\end{figure}
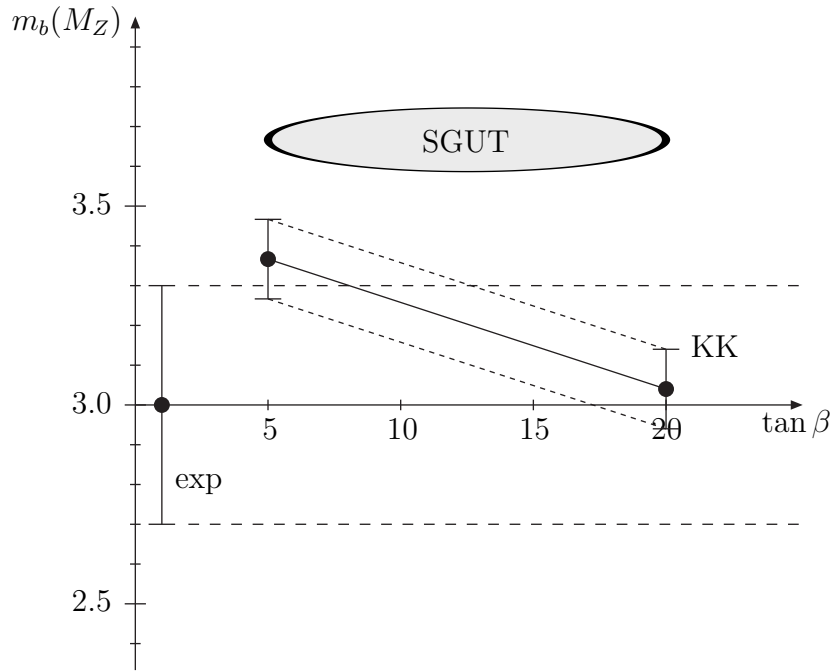
Now that the superpartner spectrum is known, the supersymmetric
threshold correction for the $b$ quark mass prediction from Yukawa
unification can be calculated. Since we are computing the correction
to a dimensionless Yukawa coupling, $\tilde{m}$ drops out and the
result depends only on $\tan \beta$: the fractional correction to the 
$b$ quark mass is $-0.006 \tan\beta$ if the $\mu$ parameter is negative. 
The prediction for $m_b(M_Z)$ is shown in Figure~\ref{fig:b-tau}, 
including the unified threshold corrections of Eq.~(\ref{eq:delta-mb-Mc}).
The supersymmetric threshold correction to the QCD coupling can also 
be calculated, and these predictions may be written in the compact 
form~\cite{Hall:2002ci}
\begin{eqnarray}
 \alpha_s(M_Z) 
 &=& \left( 0.1182 - 0.0030 \, \ln\frac{\tilde{m}}{400~{\rm GeV}}
   -0.0019 \, \ln\frac{M_s/M_c}{200} \right) \pm 0.003,
\\
 m_b(M_Z) 
 &=& \left( 3.26 - 0.022\, (\tan\beta - 10)
   -0.026 \, \ln\frac{M_s/M_c}{200} \right) \pm 0.1~{\rm GeV},
\end{eqnarray}
where we have also included the finite correction to $\alpha_s(M_Z)$ from 
$M_c$, calculated using dimensional regularization~\cite{Contino:2001si}.

The third generation squarks and sleptons are massless at tree level
since they reside on the boundary, while the squarks and sleptons of
the first two generations of $T$ have a tree level mass $\tilde{m}$ 
as they propagate in the bulk. This means that the $U(3)_T$ flavor 
symmetry is broken to $U(2)_T$ at tree level in the superpartner 
spectrum. The flavor changing effects triggered by this flavor symmetry 
breaking are expected to be larger than in conventional supersymmetric 
unification with gravity mediated supersymmetry breaking, where 
such flavor breaking occurs only through top quark radiative 
corrections~\cite{Hall:1985dx, Barbieri:1994pv, Hisano:1996qq}.
The signals are particularly important in the lepton sector.
By rotating to a mass eigenstate basis for charged leptons, while 
maintaining diagonal scalar mass-squared matrices, we can go to a basis 
where the lepton flavor violation appears only via a single new 
mixing matrix $W^e$ in the lepton-slepton-gaugino interaction:
\begin{eqnarray}
  {\cal L}^{\rm LFV} &=& - \left( \sqrt{2}g' e W^{e\dagger} 
      \tilde{e}^\dagger \tilde{b} + {\rm h.c.} \right),
\label{eq:lfvwe}
\end{eqnarray}
and in Higgs interactions.  Here, $\tilde{b}$ represents the $U(1)_Y$ 
gaugino and
\begin{equation}
  W^e = R^e_{23} R^e_{12} 
    = \pmatrix{
          c^e_{12}           & -s^e_{12}          & 0          \cr
          s^e_{12}\,c^e_{23} & c^e_{12}\,c^e_{23} & -s^e_{23}  \cr
          s^e_{12}\,s^e_{23} & c^e_{12}\,s^e_{23} & c^e_{23}   \cr },
\label{eq:we}
\end{equation}
where $c^e_{ij} \equiv \cos\theta^e_{ij}$ and $s^e_{ij} \equiv 
\sin\theta^e_{ij}$.  Therefore, we find a remarkable result that 
all the lepton flavor violating processes are completely described 
by two angles, $\theta^e_{12}$ and $\theta^e_{23}$, as far as 
the charged lepton sector is concerned. 

\begin{figure}
\begin{center}
\begin{picture}(240,85)(-120,-15)
  \ArrowLine(-120,0)(-90,0) \Text(-110,-5)[t]{$\mu_L$}
  \ArrowLine(-60,0)(-90,0)  \Text(-70,-5)[t]{$\mu_R$}  \Vertex(-90,0){2}
  \ArrowLine(-60,0)(60,0)   \Text(0,-3)[t]{$\tilde{b}$}
  \DashArrowArc(0,0)(60,0,180){2} \Text(0,68)[b]{$\tilde{e}_{Ri}$}
  \ArrowLine(110,0)(60,0)  \Text(90,-5)[t]{$e_R$}
  \Photon(64,20)(110,50){2}{7} \Text(106,58)[b]{$\gamma$}
\end{picture}
\caption{A Feynman diagram contributing to $\mu \rightarrow e\gamma$.}
\label{fig:mu-e-gamma}
\end{center}
\end{figure}
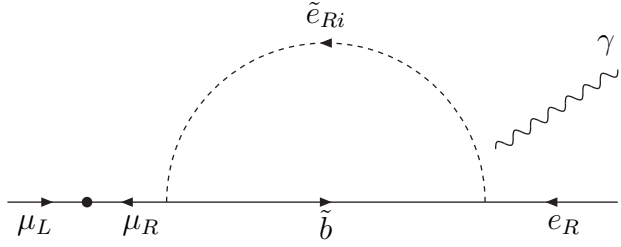
The gaugino exchange diagram for $\mu \rightarrow e \gamma$ is shown
in Figure~\ref{fig:mu-e-gamma}. Including other one-loop contributions, 
we find the branching ratio to be~\cite{Hall:2002ci}
\begin{equation}
  {\rm Br}(\mu \rightarrow e\gamma) 
    \simeq 3 \times 10^{-11} 
      \left(\frac{200~{\rm GeV}}{\tilde{m}}\right)^4
      \left(\frac{|W^e_{\tau\mu}|}{0.04}\right)^2 
      \left(\frac{|W^e_{\tau e}|}{0.01}\right)^2 
      \left(\frac{\tan\beta}{5.0}\right)^2.
\label{eq:br-meg}
\end{equation}
Here, we have normalized elements of the new mixing matrix $W^e$ by 
the corresponding values in the CKM matrix.  This is well motivated 
because $W^e$ comes from a rotation of the right-handed charged leptons 
$e$, and the rotation of $e$ is expected to be similar to that of the 
left-handed quarks $q$, which determines the CKM matrix.

The prediction given in Eq.~(\ref{eq:br-meg}) is very interesting, 
since it gives a number close to the present experimental 
bound ${\rm Br}(\mu \rightarrow e\gamma) \simlt 1.2 \times 
10^{-11}$~\cite{Brooks:1999pu}.  While we expect an uncertainty 
of a factor of a few in the estimate of Eq.~(\ref{eq:br-meg}), 
we can still say that the present $\mu \rightarrow e\gamma$ decay 
experiment has already probed the theory up to about $\tilde{m} 
\simeq 200~{\rm GeV}$ ($300~{\rm GeV}$) for $\tan\beta = 5$ ($10$). 
Furthermore, a new experiment is under construction at PSI which 
aims for a sensitivity to ${\rm Br}(\mu \rightarrow e\gamma)$ at 
the $10^{-14}$ level~\cite{PSI}. From Eq.~(\ref{eq:br-meg}), 
${\rm Br}(\mu \rightarrow e\gamma) \simlt 10^{-14}$ corresponds to 
$\tilde{m} \simgt 1.5~{\rm TeV}$ ($2~{\rm TeV}$) for $\tan\beta = 5$ 
($10$), so that this experiment will probe essentially all the 
parameter region of the theory where radiative electroweak symmetry 
breaking occurs naturally.

Another important lepton flavor violating process is $\tau \rightarrow 
\mu\gamma$ decay, which is predicted as
\begin{equation}
  {\rm Br}(\tau \rightarrow \mu\gamma) 
    \simeq 5 \times 10^{-8} 
      \left(\frac{200~{\rm GeV}}{\tilde{m}}\right)^4
      \left(\frac{|W^e_{\tau\mu}|}{0.04}\right)^2 
      \left(\frac{|W^e_{\tau\tau}|}{1.0}\right)^2 
      \left(\frac{\tan\beta}{5.0}\right)^2.
\label{eq:br-tmg}
\end{equation}
The present experimental bound comes from CLEO: ${\rm Br}(\tau 
\rightarrow \mu\gamma) \simlt 1.1 \times 10^{-6}$~\cite{Ahmed:1999gh}.
The $B$ factories at KEK and SLAC will improve the bound to the level 
of $10^{-7}$.  Note that the combination of lepton flavor violation 
mixing angles, $\theta^e_{ij}$, appearing in Eq.~(\ref{eq:br-tmg}) 
is different from that in Eq.~(\ref{eq:br-meg}).  Therefore, in 
principle, we can determine all the lepton flavor violation mixing 
angles, $\theta^e_{12}$ and $\theta^e_{23}$, by measuring both 
$\mu \rightarrow e$ and $\tau \rightarrow \mu$ transition rates, 
if we know $\tilde{m}$ and $\tan\beta$ from independent measurements 
of the superparticle spectrum. These branching ratios, together with 
the branching ratio for $\mu \rightarrow eee$ and the rate for 
$\mu \rightarrow e$ conversion, are shown~\cite{Hall:2002ci} in 
Table~\ref{table:spectrum} for two representative values of 
$\tilde{m}$ and $\tan\beta$.

\section{Conclusions}
\label{sec:conc}

We have proposed an alternative framework for physics at the scale 
of unification of the strong and electroweak interactions. The three 
pillars of physics beyond the standard model --- grand unification, 
supersymmetry and extra dimensions --- are combined in a way that 
allows calculations to a new level of precision. There are two 
keys to this framework: breaking of the unified gauge symmetry by 
local defects and the validity of the higher dimensional theory 
over a large energy range. The framework is illustrated in 
Figure~\ref{fig:machine}. The geometrical breaking of gauge symmetry 
leads to a new constrained set of theories, and a high degree of 
calculability follows because the effective theory is valid up to 
the high energy scale of strong coupling and the spectrum of KK modes 
is determined.

Local gauge symmetry breaking defects could be viewed as a step 
backwards: the Higgs mechanism provides a dynamical origin for 
symmetry breaking resulting from an underlying theory that is 
completely symmetrical. Defects represent explicit local breaking 
of gauge symmetry. They arise from the assumed form for boundary 
conditions on the fields in extra dimensions, which are presumably 
to be determined by a more fundamental theory. However, in practice 
we find that the symmetry breaking boundary conditions provide a 
simple and elegant description of nature, making the Higgs fields 
and Higgs potentials of realistic 4D grand unified theories appear 
complicated and cumbersome.

{\it The defect framework elegantly solves outstanding problems of 4D 
unification, and the simplest model fits the data with extraordinary 
accuracy.} The mass splitting of Higgs doublets and triplets is a 
necessary consequence of a bulk Higgs multiplet, proton decay from 
dimension five operators is forbidden by a spacetime symmetry of 
the bulk, and quark-lepton mass relations occur only for the heavy 
generation located on a symmetrical boundary.

The minimal model has an $SU(5)$ gauge symmetry in a 5D bulk, valid 
over the energy interval $M_c \simeq 5 \times 10^{14}~{\rm GeV}$ 
to $M_s \simeq 1 \times 10^{17}~{\rm GeV}$, as shown in 
Figure~\ref{fig:eff-theory}. The single 3-2-1 point defect 
leads to a revised picture of gauge coupling unification, illustrated 
in Figures~\ref{fig:frame} and \ref{fig:alphas_2}. Since the physics 
between $M_c$ and $M_s$ is known, unified threshold corrections can 
be computed, yielding a successful prediction for the weak mixing 
angle of extraordinary precision: $\sin^2\theta_w = 0.2313 \pm 0.0004$. 
Proton decay by $X$ gauge boson exchange is governed by the scale 
$M_c$, and a lifetime of order $10^{34}~{\rm years}$ is expected for 
the mode $e^+ \pi^0$. Predictions for other modes can also be made.

Further predictions depend on how supersymmetry is broken, and we 
have explored the consequences of having this breaking also follow 
from boundary defects in the same extra dimension.  There are two 
supersymmetries in 5D, but the boundaries of the fifth dimension can 
support only a single supersymmetry.  If there is a misalignment of 
the supersymmetries at the two boundaries by a small angle $\alpha$, 
as shown in Figures~\ref{fig:SUSY-Br}, supersymmetry is broken in the 
low energy 4D theory by an amount $\tilde{m} = \alpha M_c$. The entire 
superpartner spectrum can then be predicted in terms of $\tilde{m}$ 
and $\tan \beta$, as shown in Table~\ref{table:spectrum}. Furthermore, 
large tree-level lepton flavor violation is expected, leading to 
observable rates for $\mu \rightarrow e$ and $\tau \rightarrow \mu$ 
transitions. Finally, the supersymmetric threshold corrections to the 
quark-lepton mass relation $m_b/m_\tau$ can be computed. In the minimal 
5D model this result is more successful than in the case of 4D 
unification, and is shown in Figure~\ref{fig:b-tau}.

\section*{Acknowledgments}

This work was supported in part by the Director, Office of Science, 
Office of High Energy and Nuclear Physics, of the U.S. Department of 
Energy under Contract DE-AC03-76SF00098, and in part by the National 
Science Foundation under grant PHY-00-98840.

\end{document}